\documentclass[a4paper,fleqn,usenatbib,useAMS]{mnras}

\usepackage{graphicx}	
\usepackage{amsmath}	
\usepackage{amssymb}	
\usepackage{multicol}        
\usepackage{bm}		
\usepackage{pdflscape}	
\usepackage{epsf}
\usepackage{tikz}
\usetikzlibrary{decorations.pathreplacing, shapes, arrows}
\usetikzlibrary{fadings}

\newcommand{\kms}{\;{\rm km}\,{\rm s}^{-1}}
\newcommand{\kmsmpc}{\kms\;{\rm Mpc}^{-1}}

\newcommand{\HI}{{\sc Hi}}

\newcommand{\hmpc}{h^{-1}{\rm Mpc}}

\newcommand{\msolar}{\;{\rm M}_{\odot}}

\newcommand{\muf}{{\sc Mufasa}}
\newcommand{\mhi}{M$_\textsc{Hi}$}
\newcommand{\hir}{\mhi/M$_*$}
\newcommand{\neuron}[2]{\overset{{#1}}{y}_{#2}}
\makeatletter
\newcommand{\Rom}[1]{\expandafter\@slowromancap\romannumeral #1@}
\makeatother

\usepackage[T1]{fontenc}
\usepackage{ae,aecompl}

\usepackage{newtxtext,newtxmath}

\title[\HI~ prediction]{Predicting the Neutral Hydrogen Content of Galaxies From Optical Data Using Machine Learning}

\author[Rafieferantsoa, Andrianomena \& Dav\'e]{Mika Rafieferantsoa$^{1,2,3}$
\thanks{e-mail: \href{mailto:rafieferantsoamika@gmail.com}
{rafieferantsoamika@gmail.com}},
Sambatra Andrianomena$^{4,1}$
\thanks{e-mail: \href{mailto:andrianomena@gmail.com}
{andrianomena@gmail.com}},
Romeel Dav\'e$^{5,1,3}$
\\
\\$^1$ University of the Western Cape, Bellville, Cape Town 7535,
South Africa
\\$^2$ Max-Planck-Instit\"ut f\"ur  Astrophysik, Garching, Germany
\\$^3$ South African Astronomical Observatory, Observatory,
Cape Town 7925, South Africa
\\$^4$ SKA South Africa, 3$^\mathrm{rd}$ Floor, The Park, Park Road,
Pinelands, 7405, South Africa
\\$^5$ Institute for Astronomy, Royal Observatory, Edinburgh EH9 3HJ, UK
}

\date{Last updated 2018 March 22; in original form 2018 March 22}

\pubyear{2018}

\begin{document}
\label{firstpage}
\pagerange{\pageref{firstpage}--\pageref{lastpage}}
\maketitle

 \begin{abstract}
We develop a machine learning-based framework to predict the \HI~
content of galaxies using more straightforwardly observable quantities
such as optical photometry and environmental parameters. We train
the algorithm on $z=0-2$ outputs from the \muf\ cosmological
hydrodynamic simulation, which includes star formation, feedback,
and a heuristic model to quench massive galaxies that yields a
reasonable match to a range of survey data including \HI.  We employ
a variety of machine learning methods (regressors), and quantify
their performance using the root mean square error ({\sc rmse}) and the
Pearson correlation coefficient (\textbf{r}).  Considering SDSS photometry,
3\textsuperscript{rd} nearest neighbor environment and line of sight peculiar
velocities as features, we obtain \textbf{r} $> 0.8$ accuracy of the \HI-richness
prediction, corresponding to {\sc rmse}$<0.3$.  Adding near-IR photometry
to the features yields some improvement to the prediction.
Compared to all the regressors,
random forest shows the best performance, with \textbf{r} $>0.9$ at $z=0$,
followed by a Deep Neural Network with \textbf{r} $>0.85$.  All regressors
exhibit a declining performance with increasing redshift, which
limits the utility of this approach to $z\la 1$, and they tend to
somewhat over-predict the \HI\ content of low-\HI\ galaxies which
might be due to Eddington bias in the training sample.  We test our
approach on the RESOLVE survey data.  Training on a subset of
RESOLVE, we find that our machine learning method can reasonably
well predict the \HI-richness of the remaining RESOLVE data, with
{\sc rmse}$\sim0.28$.  When we train on mock data from \muf\ and test
on RESOLVE, this increases to {\sc rmse}$\sim0.45$.
Our method will be useful for making galaxy-by-galaxy survey predictions
and incompleteness corrections for upcoming \HI\ 21cm surveys such
as the LADUMA and MIGHTEE surveys on MeerKAT, over regions where
photometry is already available.

\end{abstract}




\section{Introduction}

\label{intro}
\indent

One of the most important science goals of the Square Kilometre
Array (SKA) project is to provide us more insights into the growth
and fueling of galaxies. A particular focus is on the evolution of
their atomic neutral hydrogen, or \HI~content, which constitutes
a major part of the gas content of galaxies, as traced by 21cm radio
emission.  \HI\ gas represents the dense gas reservoir that will
eventually form stars after passing through a molecular phase, and
is thus a key and so far underexplored aspect of the baryon cycle
governing galaxy evolution~\citep{Somerville-15}.  Hence upcoming
surveys with SKA precursors MeerKAT and ASKAP aim to expand the
depth and area of 21cm surveys out to $z\sim 1$, with the SKA
potentially reaching even higher redshifts.

Much work has been done on studying the \HI\ content of galaxies
in the nearby universe. The Arecibo Legacy Fast ALFA
\citep[ALFALFA;][]{Giovanelli-05-0} blindly observed about 7000
deg$^2$ of the Arecibo sky and was complete in 2012. It has enabled
a precise study of the distribution of galaxies in the local universe
based on their \HI~mass.  For instance, \citet{Jones-16} studied
the environmental effects on the \HI~content of galaxies using the
Arecibo Legacy Fast ALFA survey $\alpha.70$ (70\% of the final
data). They found a shift of the Schechter function knee towards
higher value in higher density environments.  Due to ALFALFA's high
positional accuracy of $< 20$ arcsec, they could explore the optical
counterparts and extend the understanding of the stellar mass growth
based on \HI~content.  The GALEX Arecibo SDSS Survey
\citep[GASS;][]{Catinella-10} used a complementary approach by
selecting $\sim800~L^*$ galaxies from the Sloan Digital Sky
Survey \citep[SDSS;][]{York-00} and observed their \HI-line spectra
until detection. \citet{Catinella-10} found that the \textit{detected}
(60\% of the 20\% observed)
\HI~richness (\hir) does not go below 40\% even for the highest
stellar masses explored ($\sim10^{11}\msolar$).
Using the full GASS dataset, \citet{Catinella-13} found an environment
dependance of the gas fraction, such that galaxies in higher host
halo masses have lower \HI~than those in less dense environments,
confirming the idea that galaxy gas content and environment are
tightly connected.  The REsolved Spectroscopy Of a Local VolumE
\citep[RESOLVE;] []{Kannappan-11} survey adopted yet another approach
by observing $\sim1500$ galaxies with ranges of stellar and gas
masses within a volume-limited $53,000$ Mpc$^3$ in the nearby
Universe.  \citet{Stark-16} used the RESOLVE data, targeting an
area within the SDSS redshift survey, and found that decreasing
\HI~richness in galaxies is related to increasing host halo mass
for a given stellar content.  These data set the stage for 
explorations to lower masses and higher redshifts to be
achieved with next-generation surveys.

Theoretical studies on the evolution of \HI~content of galaxies
have also been expanding. \citet{Cunnama-14} predicted from the
Galaxies-Intergalactic Medium Interaction Calculation ({\sc gimic})
suites of hydrodynamical simulations \citep{Crain-09}, a tight
dependence of galaxies' \HI~column density and environment:
Galaxies in groups possess extended \HI~radial profiles compared to
field galaxies. The extended radial profiles originate
from the ram pressure redistribution which they found
to dominate over the gravitational restoring forces.
Although their findings are physically grounded,
disentangling such processes remain a challenge for observers.
Related results were found using a different galaxy
formation model from \citet{Dave-13}, where \citet{Rafieferantsoa-15}
found a faster depletion of \HI~content once galaxies fall in a
more massive haloes. The specific star formation rate of those
galaxies also decreases but at rate less than that of the \HI,
indicating gas stripping from the outskirt of the galaxies inward.
\citet{Quilis-17} studied the effects of ram pressure stripping.
They used a cosmological simulated box to select a sample of galaxies
residing in clusters to do their analysis.
They found that galaxies below $10^{10}\msolar$
in stellar mass are often located at the outskirts of the clusters and have high
eccentricity.  Their interactions with the environment are more
violent resulting in faster change of the gas contents and morphologies
of the galaxies.  More massive galaxies are situated closer to the
cluster centers, and the gas removal is less dominant. The major
change in those galaxies is caused by inflowing gas from the
intercluster medium.
Using the \muf~data \citep{Dave-16}, \citet{Rafieferantsoa-18}
found a weak but extended galactic conformity in \HI~richness
for galaxy members of low-mass haloes. Bigger host-halo galaxies
tend to have stronger but less extended conformity.
These studies demonstrate that the \HI\ content
of galaxies is impacted by their environment, but the exact nature
of that dependence is not entirely clear.

Hence observational surveys suggest that understanding the baryon
cycle requires precise measurements of the \HI~content of the
galaxies, which at times might be affected by observational artifacts.
Theoretical works, on the other hand, predict physical results that
are currently difficult to observe, which argues for larger and
deeper \HI~surveys to improve our current understanding of the
evolution of gas content and hence galaxy growth overall.

Although considerable efforts have gone into studying the gas phase
properties of galaxies with the help of the currently available
\HI~data, \textit{e.g.} ALFALFA and RESOLVE, the understanding of \HI\
evolution still lags behind the understanding of their stellar
components. The main reason is that photometric data can be directly
related to the stellar population of galaxies, and such optical and
near-infrared data is currently technologically able to reach
deeper levels than radio data.  For the promise of multi-wavelengths
surveys to be fully realised into the radio regime, it is important
to be able to relate gas and stellar properties accurately.  However,
this is not straightforward.  There have been attempts that have been
proposed to estimate gas-phase properties of galaxies from their
stellar masses obtained from spectral energy distributions (SED)
fitting to photometrical properties. For instance, \citet{Kannappan-04}
found a correlation between $u-K$ colours and \HI~richness which
they dubbed {\it photometric gas fractions}. The correlation was
shown to be valid for galaxies with atomic gas fraction ranging
from 1\% to 10$\times$ the stellar masses. \citet{Zhang-09} developed
a similar method by using the $i$-band surface brightness and the
$g-r$ colour to estimate the \HI~richness of the galaxies. They
found a tighter scatter compared to previous estimations.  The
\HI~scaling relations found by \citet{Zhang-09} were further improved
upon by \cite{Wang-2013} by introducing a form of correction to
account for the fact that \HI~rich galaxies have more active star
formation on the outer discs (bluer) \citep[see ][]{Wang-2011}. Still
with the standard approach by first establishing correlation between
the gas fraction and other galaxy properties, \citet{Catinella-10}
prescribed another relation $\log_{10}\rm{(M_{HI}/M_{*})} = -0.332
\log_{10}(\mu_{*})-0.240(NUV-r)+2.856$ which was also tested
by \citet{Wang-2015} with their samples to estimate the gas fraction
as a function of stellar mass surface density ($\mu_{*}$) and
observed $NUV - r$ colour.  From these studies it is clear that developing ways
to connect optical/NIR information with \HI\ is an important task,
which affords many applications such as to estimate the \HI~content
of certain galaxies based solely on their available photometry
information, to enable larger statistics, and to assess incompleteness
in surveys.

In this work, we propose a more general approach compared to previous
studies by investigating the feasibility of predicting the \HI~
richness of galaxies from the available optical properties of
galaxies, particularly the photometric magnitudes and environmental
quantities, using machine learning.  The main idea is that machine
learning can synthesise all the photometric data in order to optimally
predict \HI, rather than trying to isolate particular combinations
that work best.  The advantage of using machine learning techniques
is mainly the capability of the model to learn peculiar aspects
human might have overlooked, with the downside that such a method
does not provide a direct physical interpretation of the result.
By using simulated galaxies to train and calibrate the method,
connections can be made between the obtained correlations and the
underlying physics, at least within the context of the given model.
In this paper, the first in a series, we focus on galaxies having
at least some \HI~content; future works will consider identifying
gas-free galaxies.  Our best machine learning algorithms, random
forest and deep learning, are able to predict the \HI~richness of
simulated galaxies to within $<0.3$ dex from their real values using
only the photometric properties of the simulated galaxies. Testing
this on the RESOLVE survey, the prediction of the observed data
from simulation-trained models yield less precise results. Generally,
random forest is our optimal machine learning algorithm, but the
neural network's performance becomes better when observational data
are used.

Our method has numerous applications.  Current data as well as
future surveys will benefit from this method by providing ways to
more accurately correct observations for incompleteness and confusion.
For instance, the upcoming Looking At the Distant Universe with the
MeerKAT Array \citep[LADUMA;][]{Holwerda-12} survey aims to directly
detect and use different techniques to stack multiple objects to
be able to measure \HI~fluxes out to $z>1$ for the first time, to
enable a deeper understanding of the fueling processes of galaxies
and study the cosmic evolution of their \HI~content.  But at higher
redshift, confusion can become dominant especially when sources are
located in groups.  Meanwhile, ASKAP \HI~All-Sky Survey (WALLABY)
which will cover two third of the sky will probe \HI~gas of $6\times
10^5$ galaxies up to $z = 0.26$; DINGO, up to $z = 0.43$, will probe
about $10^5$ galaxies within $4\times 10^7\>\rm{Mpc}^{3}$ cosmological
volume \citep{Duffy-12}.  These \HI~surveys will provide a wealth
of information on galaxy evolution, but it is important to be able
to accurately measure and understand the observations, which is
where our method can provide insights.

\S\ref{sim} briefly reviews the \muf\ simulation used for this work.
The approach we use in this study is detailed in \S\ref{S:methods},
and we present the techniques utilized in order to achieve our goal
in \S\ref{S:models}. \S\ref{results} presents our findings and
\S\ref{S:test} shows a preliminary application of our method.  We
expand on the limitations of our method in \S\ref{weakness} and
finally conclude in \S\ref{conclusion}.

\section{Simulations}\label{sim}
\subsection{Galaxy formation models: \muf}\label{model}
\indent

For our training set we make use of the outputs of the \muf\
simulation model, which is fully described in \citet{Dave-16}.  We
only present the key prescriptions in the model that are particularly
relevant for this work.

\muf\ is implemented in the {\sc Gizmo} cosmological hydrodynamics
code, including a tree-particle-mesh gravity code based on {\sc
Gadget}~\citep{Springel-05}, topped with a meshless finite mass
hydrodynamic algorithm \citep{Hopkins-15}.  The model uses radiative
cooling and heating implemented with the {\sc Grackle 2.1}
library\footnote{\url{https://grackle.readthedocs.io/en/grackle-2.1/genindex.html}}.
Star formation follows a \citet{Schmidt-59} law, based on a subgrid
prescription that determines the molecular gas content of each gas
particles~\citep{Krumholz-Gnedin-11}, and occurs only in gas elements above
a hydrogen number density threshold of $n_H>0.13$cm$^{-3}$.

\muf\ uses a kinetic gas outflow prescription to model star-formation
driven winds, following scalings from the Feedback in Realistic
Environments~\citep[FIRE;][]{Muratov-15} zoom simulations.  \muf~also
contains a heuristic prescription for star formation quenching
whereby it heats the gas volume elements within a host halo that
are above a halo mass threshold of $M_{\rm halo}>(1+0.48z)\times
10^{12}M_\odot$~\citep{Gabor-Dave-15,Mitra-15}.  This model is intended
to mimic radio mode feedback from active galactic nuclei
\citep{Croton-06} in massive halos.

\subsection{Galaxy sample}\label{sample}

The galaxy sample used for our analysis is obtained by simulating
a cube of $50\hmpc$ on a side with $512^3$ dark matter particles
and $512^3$ gas volume elements. The initial conditions are generated
at redshift $z=249$ using {\sc Music} \citep{Hahn-11} with
\citet{-16}-concordant cosmological parameters, namely $\Omega_m =
0.3$, $\Omega_\Lambda = 0.7$, $\Omega_b = 0.048$, $H_0 = 68 \kmsmpc$,
$\sigma_8 = 0.82$ and $n_s = 0.97$.

\muf\ evolves these initial conditions to $z=0$, outputting 135
snapshots.  For each snapshot, we identify galaxies, with {\sc
Skid}\footnote{\url{http://www-hpcc.astro.washington.edu/tools/skid.html}
\label{skid}}~\citep{Keres-05} as gravitationally bound collections
of stars and star-forming gas. In our analysis, we will only use
$z\leq2$ sample, which, in total, is made of 50 snapshots.  Each
snapshot contain typically around 8000 resolved galaxies ($>64$
star particle masses or $M_*>1.16\times 10^9M_\odot$).

\subsection{Galaxy properties}
Our simulated galaxy properties are calculated with a modified version of {\sc
caesar}\footnote{\url{https://bitbucket.org/laskalam/caesar}}, which
is an add-on package for the {\sc yt} simulation analysis suite.
The stellar mass of a galaxy, or M$_*$, is the total mass
of the stellar particles within it. 
The atomic neutral hydrogen content,
\mhi, of the galaxy is the summation of all \HI~from the gas particles. 
For each gas volume element, we account for the self-shielding
from the metagalactic UV background radiation, by using a fitting formula
for the effective optically-thin photoionization rate as a function of density~\citep{Rahmati-13}.
The galaxy peculiar velocity v$_\mathrm{gal}$ is the 1-D mass-weighted average
of all the particle velocities contained in it, along each of the $(x,y,z))$ directions.
We use the projected nearest neighbour density $\Sigma_3$
to quantify the galaxy environment such that:
\begin{equation}
\Sigma_3 = {3\over \pi R_3^2}
\end{equation}
where $R_3$ is the distance of the galaxy to its 3rd closest 
neighbour, projected along the line of sight.

The magnitudes of the galaxies are obtained using the {\sc
Loser}\footnote{Line Of Sight Extinction by Ray-tracing\\
\url{https://bitbucket.org/romeeld/closer}} \citep[see][for a fuller
description]{Dave-17b} package (not {\sc caesar}) but still using
the groups identified by {\sc Skid}.  We first use the Flexible
Stellar Population Synthesis \citep[FSPS;][]{Conroy-10} library to
derive the stellar spectra of each star particle based on its age
and metallicity, summing to obtain the stellar spectrum for that
galaxy. Every stellar spectrum is attenuated by the line of sight
dust extinction obtained by scaling the metal column density along
the given line of sight; this results in each of 6 lines of
sight $(\pm x,\pm y,\pm z)$ having different extinction and thus
different spectra.  We obtain all magnitudes by applying the
appropriate filters.  We computed (\textit{u,g,r,i,z}) SDSS magnitudes,
(\textit{U,V}) Johnson magnitudes, \textit{NUV} GALEX magnitude,
and the (\textit{J,H,K$_s$}) 2MASS magnitudes.

\section{Machine Learning Setup} \label{S:methods}

The goal is to predict the \HI~richness (\hir) from other properties
of a given galaxy. We use the supervised learning paradigm
which consists of training the algorithm to estimate the desired label
when fed with a corresponding input.
Through a learning process, the best model parameters that minimize
a defined cost function,  which we choose to be the mean squared errors
({\sc mse}), are solved. Sets of training datasets drawn from
our simulated sample are used to train our learners to predict the target (\hir)
from the features $\{u,g,r,i,z,U,V,J,H,K_s, \Sigma_3, v_{gal}\}$ of our galaxies. 

It is noted that $v_{gal}$ indicates line of sight velocity,
and our models will predict the \HI~richness (\hir)
of the galaxies rather than their \mhi~due to the less constrained correlation
between the latter and the galaxy stellar masses.
In addition, we take the logarithmic values of the target due to
its large dynamic range which can cause the learning process to fail.
First of its series, this work focuses only on the prediction of
the \HI~richness of \HI~rich galaxies and to do so, we only select
galaxies with \hir$>10^{-2}$, which decreases the size
of our sample. To counteract, we increase our data by calculating
the galaxy properties along all the 6 projections axis of the simulated cubical box,
resulting in $6\times$ more data for our analysis.

We assume we have all photometric magnitudes for all available bands,
covering a wide range of spectrum including SDSS magnitudes,
Johnson magnitudes and 2MASS magnitudes, which we can compute from
our simulated galaxies. Although this scenario is ideal for our analysis,
it is not so realistic. We can expect observed galaxies to only have \{{\it u,g,r,i,z}\}
magnitudes at best. To this regard, we examine different possibilities in our
analysis. All the setups considered in this work are listed in Table \ref{setups},
where \verb'color indices' denotes all possible pairwise combination (\textit{e.g.}
$g-r$) of all the magnitudes in the surveys considered in one setup.

\begin{table*}
 \caption{List of all the setups that are considered in the analysis.
 For easy reference, each setup has been given a name.}
 \label{setups}
 \begin{tabular}{lllll}
  \hline
  Name & Surveys & Features & Target & Description\\
  \hline
  \hline
  fSMg  & SDSS & $u,\> g,\> r,\> i,\> z,\> v_{gal},\> \Sigma_{3}$
            & $\rm log(M_{HI}/M_{*})$ & redshift information not required\\[2pt]
  fSClr  & SDSS & \verb'color indices',$\>v_{gal},\> \Sigma_{3}$
            & $\rm log(M_{HI}/M_{*})$ & redshift information not required\\[2pt]
   fSCmb & SDSS
               & \verb'color indices', $u,\> g,\> r,\> i,\> z,\> v_{gal},\> \Sigma_{3}$
               & $\rm log(M_{HI}/M_{*})$ & redshift information not required\\[2pt]
  fAMg    & SDSS+Johnson+2MASS
              & $H,\>J,\> Ks,\> U,\> V,\>u,\> g,\> r,\> i,\> z,\>v_{gal},\> \Sigma_{3}$
               & $\rm log(M_{HI}/M_{*})$ & redshift information not required \\[2pt]
  fAClr  & SDSS+Johnson+2MASS
            & \verb'color indices',$\>v_{gal},\> \Sigma_{3}$ & $\rm log(M_{HI}/M_{*})$    
            & redshift information not required \\[2pt]
  \hline
  zSMg & SDSS & $u,\> g,\> r,\> i,\> z,\> v_{gal},\> \Sigma_{3}$
            & $\rm log(M_{HI}/M_{*})$ & prediction at a given redshift bin\\[2pt]
  zSClr & SDSS & \verb'color indices',$\>v_{gal},\> \Sigma_{3}$
            & $\rm log(M_{HI}/M_{*})$ & prediction at a given redshift bin\\[2pt]
  zSCmb & SDSS
               & \verb'color indices', $u,\> g,\> r,\> i,\> z,\> v_{gal},\> \Sigma_{3}$
               & $\rm log(M_{HI}/M_{*})$ & prediction at a given redshift bin \\[2pt]
  zAMg & SDSS+Johnson+2MASS
            &$H,\>J,\> Ks,\> U,\> V,\>u,\> r,\> r,\> i,\> z,\>v_{gal},\> \Sigma_{3}$
            & $\rm log(M_{HI}/M_{*})$ &prediction at a given redshift bin \\[2pt]
  zAClr & SDSS+Johnson+2MASS
            & \verb'color indices',$\>v_{gal},\> \Sigma_{3}$
            & $\rm log(M_{HI}/M_{*})$ & prediction at a given redshift bin \\[2pt]

  \hline
 \end{tabular}
\end{table*}

We train our model in two different ways. First is the ``$f$-training'',
which considers all the galaxies from all the $z\leq2$ outputs
(with $f$ leading the setup names, see first column of Table \ref{setups}).
Second is the ``$z$-training'', in which we build a regressor at each redshift bin
(with $z$ leading the setup names).
In both approaches, we randomly choose 75\% of the data as the training set
and 25\% as testing set. We do the training 10 times with 10 different random
batches to get the uncertainty of our results\footnote{At each iteration,
the dataset is randomly shuffled and new batches of training
and test sets are generated.}.

To this end, we make use of 6 different machine learning techniques
that we describe in the following.

\section{Machine Learning Algorithms}\label{S:models}

We use {\tt TensorFlow} to build the DNN model and {\tt scikit-learn} \citep{scikit-learn} package for the remaining methods. 
\subsection{Linear regression (LR)}
Linear regression model (along with kNN, see \S\ref{knn}) is
the simplest amongst those we use in this work.
Its simplicity, hence its great speed during training, provides quick insights
into the relationship between the features ({\bf x}) and the corresponding
target ($y$). The latter is defined as a linear combination of all the features,
$y = \textbf{w}\cdot \textbf{x}$, and the idea consists of finding the weights
$\textbf{w}$ that minimize the mean squared error ({\sc mse})
\begin{equation}
\textsc{mse} = 
     \frac{1}{N}\sum_{n=1}^{N}(\textbf{w}\cdot \textbf{x}_{n} - y_{n})^{2}.
\end{equation}
Here the bias is absorbed into the weights $\textbf{w}$.

\subsection{Ensemble learning methods: Random forest (RF)
and Gradient Boosting (GRAD)}
To understand both RF and GRAD algorithms one needs to first look
at their base estimators, the Decision trees \citep{Hastie-09}, which
will be clarified bellow.

In a simple one dimensional problem, we assume a dataset
$\mathcal{D}$ = $\{(x_{1}, y_{1}),(x_{2}, y_{2}),...,(x_{N}, y_{N})\}$ of length $N$
($(x,y) \in \mathbb{R} \times \mathbb{R}$). The first step of the algorithm is
to split the training set at a split point $s$ that minimizes the cost function

\begin{equation}\label{cost-tree}
J = \underset{c_{1}}{\rm{min}} \left\{\sum_{x_{i} \in R_{1}(s)}(y_{i}-c_{1})^{2}\right\} +
\underset{c_{2}}{\rm{min}} \left\{\sum_{x_{i} \in R_{2}(s)}(y_{i}-c_{2})^{2}\right\}, 
\end{equation}
where $R_{1} = \{x_{i}|x_{i} \leq s\}$ and $R_{2} = \{x_{i}|x_{i} > s\}$ are
the two regions (also called nodes) resulting from the split.
The values $c_{1}$ and $c_{2}$ that minimize each term in Eq.~\ref{cost-tree} are
simply the averages of the labels $y_{i}$ in $R_{1}$ and $R_{2}$ respectively; $i.e.$

\begin{eqnarray}\label{pred-node}
c_{1} = \frac{1}{m_1}\sum_{x_{i} \in R_{1}(s)} y_{i}, \nonumber \\
c_{2} = \frac{1}{m_2}\sum_{x_{i} \in R_{2}(s)} y_{i}, 
\end{eqnarray}  
where $m_1$ and $m_2$ are the number of inputs $x_{i}$ found in $R_{1}$ and
$R_{2}$ respectively. To grow the tree, each resulting node from the root
is further split recursively (known as greedy algorithm) until a fixed maximum depth
(or size) of the tree is reached. The nodes at the bottom of the tree are called
the leaf nodes. To predict a new label $y_{\rm new}$ from a new input $x_{\rm new}$,
one simply walks through the tree from the root to reach a leaf node which then
estimates $y_{\rm new}$ by averaging the corresponding labels $y_{i}$ of the inputs
$x_{i}$ whithin it according to\footnote{Similar to Eq.~\ref{pred-node}.}
\begin{equation}
\hat{y}_{\rm new} = \frac{1}{m}\sum_{x_{i} \in \mathcal{L}} y_{i},
\end{equation}
where $\mathcal{L}$ indicates the leaf node and $m$ the number of
points $x_{i}$ within it. Decision trees are prone to overfitting but there exist
various techniques of regularization.

Random forest \citep{Breiman-01}, known to be a powerful machine learning algorithm,
is composed of a given number\footnote{Which is among the hyper-parameters
of the model.} of decision trees (base estimators) which are individually trained with
a random subset of the dataset. To do a prediction, RF simply averages
the predictions of its decision trees.

Another well known ensemble learning model that we use is
gradient boosting \citep{Friedman-00}. Its base learner is also a decision tree but 
instead of simply aggregating the predictions of its regressors like in the case of RF,
the training is carried out in a sequence. Except for the first regressor, which is trained
with the dataset, each next regressor in the sequence\footnote{This is set by
the number of the base estimators.} fits the residual errors of its predecessor and so on.
The resulting estimator, is then of the following form
\begin{equation}
\mathcal{E}(x) = \mathcal{E}_{1}(x) + \sum_{i = 2}^{N}\gamma_{i}e_{i}(\epsilon_{i}),
\end{equation}
where $\mathcal{E}_{1}(x)$ is the first estimator, $\epsilon_{i}$ the residual errors from
the $i-1^\mathrm{th}$ learner used as inputs of the $i^\mathrm{th}$ learner to fit a predictor $e_{i}$
and $\gamma_{i}$ is a coupling parameter which is optimized such that the error from
the combined system at each iteration ($i.e.\>\mathcal{E}_{i+1}(x) = \mathcal{E}_{1}(x) +
\sum_{k = 2}^{i}\gamma_{k}e_{k}(\epsilon_{k})$) is minimized. $N$ is the number
of base regressors (equal to the number of iteration) that form the ensemble.

\subsection{k-Nearest Neighbor (kNN)}\label{knn}
$k$-Nearest Neighbour \citep{Altman-92} is a flexible non-parametric
regression algorithm. Considering a set of instances $\textbf{x}_{n}$
(in general $\textbf{x}_{n}\in\mathbb{R}^{d}$ but for the sake of simplicity
we let $\textbf{x}_{n}\in\mathbb{R}$) with their corresponding label
$y_{n}$ ($y_{n}\in\mathbb{R}$), to predict a new label $y_{\rm new}$
given a new instance $\textbf{x}_{\rm new}$, the estimate of $y_{\rm new}$
is simply the weighted average of targets of the $k-$closest neighbours
of $\textbf{x}_{\rm new}$. The principle is generalised for $d-$dimensions
in feature space. 

\subsection{Support Vector Machine (SVM)}
Given a set of training data consisting of examples
$\textbf{x}_{n}$ ($\textbf{x}_{n}\in\mathbb{R}^{d}$) and their labels
$y_{n}$ ($y_{n}\in\mathbb{R}$), the method aims at finding a linear function
of the form $f(\textbf{x}) = \textbf{w}\cdot\textbf{x} + b$.
This can be seen as a convex optimization which seeks to
\begin{itemize}

\item minimize $\frac{1}{2}\textbf{w}^{T}\textbf{w}$,\\
 subject to the constraint $|y_{n} - (\textbf{w}\cdot\textbf{x}_{n} + b)| \leq \epsilon$, 

\end{itemize}
where $\epsilon$ denotes the residuals between estimates and the desired outputs.
To deal with otherwise intractable optimization problem, \cite{Vapnik-95} introduced
some slack variables $\xi^{-}_{n}, \xi^{+}_{n}$ such that it now aims at

\begin{itemize}

\item minimizing $\frac{1}{2} \textbf{w}^{T}\textbf{w} + C\sum_{n=1}^{N}(\xi^{-}_{n} + \xi^{+}_{n})$\\
subject to
\begin{equation}
\rm{the~constraints} \> 
\begin{cases}
y_{n} - (\textbf{w}\cdot\textbf{x}_{n} + b) \leq \epsilon + \xi^{-}_{n}\\
\textbf{w}\cdot\textbf{x}_{n} + b - y_{n} \leq \epsilon + \xi^{+}_{n}\\
\xi^{-}_{n} , \xi^{+}_{n} \geq 0
\end{cases}
\end{equation}

\end{itemize} 
where $C$ is a positive value used for regularization. For simplicity, we only present
the linear case but to deal with non-linearities one can resort to a kernelized SVM.
It is noted that SVM method is also used for classification problem \citep{Cortes-95}. 

\subsection{Artificial neural network}
We dedicate this section for a rather extended description
of the deep neural network used  for this work. This is so due to
its novel application in astronomy. This is not so much the case with other
machine learning techniques described before, as they are at some point fully 
or partly used to analyze astronomical data.

\begin{figure}
	\centering
	\begin{tikzpicture}[shorten >=1pt]
		\tikzstyle{unit}=[draw,shape=circle,minimum size=0.5cm]
		\tikzstyle{hidden}=[draw,shape=circle,minimum size=0.5cm]

		\node[unit](x0) at (0,3.5){H};
		\node[unit](x1) at (0,2){Ks};
		\node at (0,1){\vdots};
		\node[unit](xd) at (0,0){v$_\mathrm{gal}$};
		\node[hidden](h10) at (2.5,4){$_{\Large y_0}^{\small 0}$};
		\node[hidden](h11) at (2.5,2.875){$_{\Large y_1}^{\small 0}$};
		\node[hidden](h12) at (2.5,1.75){$_{\Large y_2}^{\small 0}$};
		\node (h13) at (2.5,0.625){\vdots};
		\node[hidden](h1m) at (2.5,-0.5){$_{\Large y_\mathrm{m}}^{\small 0}$};
		\node[hidden](h20) at (3.5,4){$_{\Large y_0}^{\small 1}$};
		\node[hidden](h21) at (3.5,2.875){ $_{\Large y_1}^{\small 1}$};
		\node(h22) at (3.5,1.685){$\vdots$};
                \node[hidden](h2m) at (3.5,-0.5){$_{\Large y_\mathrm{p}}^{\small 1}$};
		\node[hidden](h30) at (4.5,4){$_{\Large y_0}^{\small 2}$};
		\node(h31) at (4.5,2.3125){$\vdots$};
		\node[hidden](h32) at (4.5,0.625){ $_{\Large y_\mathrm{q-1}}^{\small 2}$};
                \node[hidden](h3m) at (4.5,-0.5){$_{\Large y_\mathrm{q}}^{\small 2}$};
                \node[unit](output) at (7, 1.75){M$_\mathrm{HI}$};
		\draw[->] (x0) -- (h11);
		\draw[->] (x0) -- (h10);
		\draw[->] (x0) -- (h1m);

		\draw[->] (x1) -- (h11);
		\draw[->] (x1) -- (h12);
		\draw[->] (x1) -- (h1m);

		\draw[->] (xd) -- (h10);
		\draw[->] (xd) -- (h1m);

                \draw[-] (h10) -- (h21); 
                \draw[-] (h10) -- (h2m);

                \draw[-] (h11) -- (h21);
                \draw[-] (h11) -- (h22);

                \draw[-] (h12) -- (h2m);

                \draw[-] (h1m) -- (h20);
                \draw[-] (h1m) -- (h2m);
               
                \draw[-] (h13) -- (h21); 
                \draw[-] (h13) -- (h22); 
                \draw[-] (h13) -- (h2m);

                \draw[-] (h20) -- (h30);
                \draw[-] (h20) -- (h32);

                \draw[-] (h21) -- (h31);

                \draw[-] (h2m) -- (h3m);

                \draw[-] (h22) -- (h30);
                \draw[-] (h22) -- (h31);
                \draw[-] (h22) -- (h32);
                \draw[-] (h22) -- (h3m);

                \draw[->] (h30)  -- (output);
                \draw[->] (h31)  -- (output);
                \draw[->] (h32)  -- (output);
                \draw[->] (h3m)  -- (output);

                \draw[-,line width=0.5mm, gray] (x1)-- (h2m);
                \draw[-,line width=0.5mm, gray] (h11)-- (h3m);
                \draw[->,line width=0.5mm, gray] (h12)-- (output);



		%
		%
		%
		%
		
		\draw [decorate,decoration={brace,amplitude=5pt},xshift=-4pt,yshift=0pt] (-0.3,4) -- (0.55,4) node [black,midway,yshift=+0.6cm]{\large$^\mathrm{Galaxy}_\mathrm{properties}$};
		\draw [decorate,decoration={brace,amplitude=5pt},xshift=-4pt,yshift=0pt] (2,4.5) -- (5,4.5) node [black,midway,yshift=+0.6cm]{hidden layers};
	\end{tikzpicture}
	\caption[Network graph for a 4-layer perceptron.]{Network graph of our 4-layer perceptron with 1 output unit. The hidden layers contain $m,p,q$ neurons respectively.}
	\label{fig:multilayer-perceptron}
\end{figure}
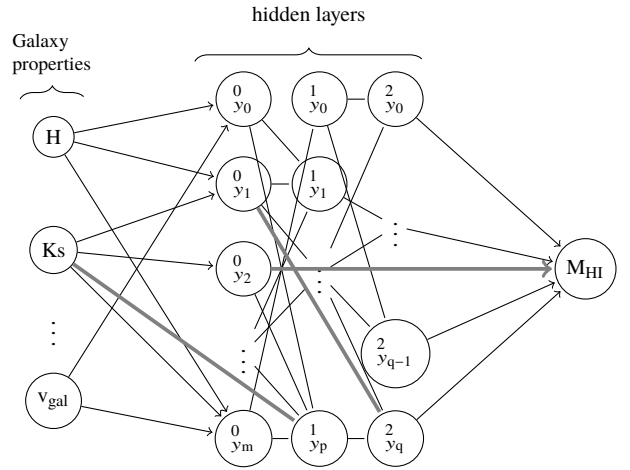

Due to our hardly correlated features and target, the choice of model to learn
the connection between them is very complex,
though our maximum number of galaxy properties are limited
to only 12 components. Figure \ref{fig:multilayer-perceptron} shows a summary of
our multilayer perceptron model.
The left nodes show our galaxies properties as input into our 3 hidden layers
and the right most node is the output.
$\neuron{j}{k}$ represents the $k^\mathrm{th}$ neuron
in the $j^\mathrm{th}$ layer and is the linear weighted sum of the preceding
neurons as shown in equation \ref{eq:neuron}, $f_a$
being the activation function (see \ref{S:activation-function}).
\begin{equation}
\label{eq:neuron}
\neuron{j}{k} = f_a\left(\sum_l w^j_{k,l}\times\neuron{j-1}{l}+ b^j_k\right)
\end{equation} 
$w^j_{k,l}$ and $b^j_k$ are the weight and bias of $\neuron{j-1}{l}$
on $\neuron{j}{l}$.

A deep neural network (DNN) is then to learn the (close to the) correct
values of $w$'s and $b$'s for the model to be able to reproduce the {\it target}
given the {\it features}.

The choices for the number of the hidden layers,
the activation functions between layers and the optimiser
are described in the following subsections.

\subsubsection{Hidden layers}
\label{S:hidden-layers}
One of the toughest step that one has to overcome in building a DNN model
is the choice of the number of hidden layers and the respective
number of neurons in each layer. The use of models with a single hidden layer
or the so called {\it universal approximators} has been advocated since
the artificial neural network was used into solving
physical problems. \citet{Cybenko-89} stated that a single hidden layer
in a feedforward\footnote{Any connections between neurons do no form a cycle}
neural network is enough to capture  the continuous non-linearity between
the inputs and the outputs. This conclusion was extended later on by
\citet{Hornik-91} that the nature of the feedforward structure drives
its universality irrespective to the activation function as long as
the latter is continuous, bounded and non-constant 
(see. \S\ref{S:activation-function}).
The ``{\it universal approximation}'' principle ended recently after the work
done by \citet{Hinton-06}. They explored the improvement of
the multi-hidden-layer architecture and concluded the following.
Although a single hidden layer with finite number of neurons can be
enough to map the connection between the input(s) and the output(s),
one extra layer is useful to increase the accuracy of the mapping.
Any additional layer is only for the model to explore possible representations
of the map and to decrease the learning time given a set of data.\\

For those reasons, and after a trial-and-error approach,
we opt to use 3 hidden layers in our model. We use 100 neurons in each layer
to correctly map the galaxy properties with all their possible combinations.
We have extra nodes to account for some degrees of freedom for safety.

\subsubsection{Activation function}
\label{S:activation-function}
Given a set of values fed to one node in our model
(see Figure \ref{fig:multilayer-perceptron}), one has to decide how much of that
information should be passed to the next connected node(s).
This can be defined with an activation function.
A sigmoid function was widely used in the past. Problems occur with that
function when the input values of a node are high (or small in the negative end):
that is the vanishingly small gradient at those ends.
In our model, we use a rectified linear unit  function ($\mathrm{RelU}$,
see eq. \ref{eq:relu}). It means that any negative values passing
the nodes are set to zero (ignored).
\begin{equation}
\label{eq:relu}
f(x) = max(0,x)
\end{equation}
We also tested the use of an exponential linear unit function
($\mathrm{elU}$, see eq. \ref{eq:elu}). In this case, we allow a small fraction
of the negative signal to go through the next connected node(s).
\begin{equation}
\label{eq:elu}
  f(x)=\begin{cases}
    x, & \text{if $x\geq0$}.\\
    \exp{(x)} -1, & \text{otherwise}.
  \end{cases}
\end{equation}
Our test didn't get any improvement (if not deterioration) in using $\mathrm{eLU}$.
Using different activation functions such as {\it hyperbolic tangent}, {\it gaussian}
or {\it multiquadratics} are not favoured in our case.

\subsubsection{Optimisation}
\label{S:optimisation}
After each step of calculations, the network should optimize the model based
on its current and previous states to improve the subsequent mapping.
Our model utilizes a computationally memory efficient optimization
due to its dependancy to only the first order gradients, namely the
``{\it adaptive moment estimation}'' (or Adam).
For more details we refer the readers to \citet{Kingma-14}.
Adam optimization, compared to other gradient-based optimization,
is very suitable for noisy and sparse gradients, and for simulated data
which show very large scatter with respect to
a given quantity of parameter \citep{Kingma-14}.
With this optimizer, we have to decide few parameters in advance.
The learning step $\alpha$ and the parameters controlling the moving averages
of the $1^\mathrm{st}$ and $2^\mathrm{nd}$ order moments namely $\beta_1$
and $\beta_2$ (both $\in$[0,1)) respetively. For this purpose,
we chose to minimize the mean squared error between the target 
and the prediction from the model: in what follows,
we will alternatively call the mean
squared error  the ``{\it objective function}'' $f$({\bf x}): with {\bf x} the parameters
of the model to be updated, such as weights and biases.
At a given time $t\leq T$, where $T$ is the maximal learning time step,
we can update the parameters of the model as shown in the following.
 
\begin{align}
g_t                      &=\nabla_\mathrm{ x} f(\mathrm{x}_{t-1})\label{eq1}\\
 \mu_{1,t}           &=\beta_1 \times \mu_{1,t-1} + (1-\beta_1)\times g_t\label{eq2}\\
 \bar{\mu}_{1,t}  &=\mu_{1,t}/(1-\beta_1^t)\label{eq3}\\
 \mu_{2,t}           &=\beta_2 \times \mu_{2,t-1} +
                               (1-\beta_2)\times g_t^2\label{eq4}\\
 \bar{\mu}_{2,t}  &=\mu_{2,t}/(1-\beta_2^t)\label{eq5}\\
 \mathrm{x}_t     &=\mathrm{x}_{t-1} - \alpha_t \times \bar{\mu}_{1,t}/
                               (\sqrt{\bar{\mu}_{2,t}} + \epsilon) \label{eq6}
\end{align}
where $\alpha_t=\alpha\sqrt{1-\beta_2^t}/(1-\beta_1^t)$ is the time-step at $t$.
Equation \ref{eq1} shows the gradients of the objective function at $t$ with respect
to the model parameters. Equations \ref{eq2},\ref{eq4} update the estimations
of the 1$^\mathrm{st}$ and 2$^\mathrm{nd}$ moments. Our moments are 
biased towards the initial values, thus we require equations \ref{eq3},\ref{eq5}
to account for  the corrections.
Finally, we update the model parameters with equation \ref{eq6}.

We do not claim that the choice of parameters implemented in our models
as well as their configurations are the best to do similar work.
We will likely continue to improve this method in subsequent papers.
\section{\HI\ Prediction Using Machine Learning}\label{results}

Our goal is to predict the \HI\ richness of a given galaxy based
on its optical/near-IR photometry.  We choose to predict \HI\ richness and not
\HI\ mass as it is expected to correlate more with galaxy
colours, with \HI-poor galaxies being redder than \HI-rich ones,
so in some sense gives more physical information than just \HI\
mass alone which approximately correlates with stellar mass.
Nonetheless, our approach could equivalently be used for either,
and we have tested that the resulting accuracy of the predictions
is similar.

\subsection{Quantifying the mapping accuracy}
\begin{figure}
\includegraphics{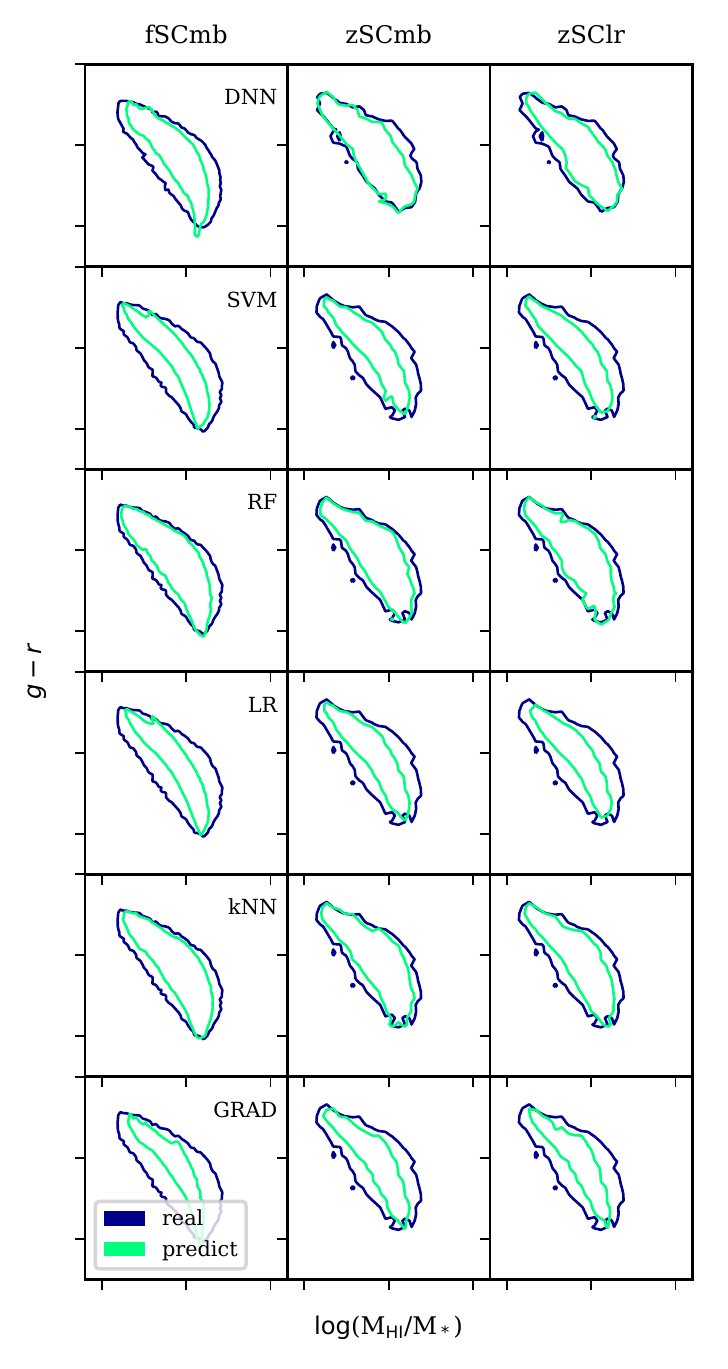}
\caption{Superposition of the predicted (green) and the real (blue) \HI~richness
of our galaxies (x-axes) \textit{vs.} $g-r$ colour (y-axes). The contours are enclosing $2\sigma$
of the distributions. Each row shows different mapping corresponding
a particular method and each column a different setup (see Table \ref{setups}).}
\label{fig_2D_superp}
\end{figure}
For a given trained model (see \S\ref{S:methods}), we can predict
the \HI~richness of a test set which contains the feature
parameters, similar to those used during the training, and the real \HI~richness.
One can then see for a given example (composed by the features) how the model estimates the corresponding \HI~richness and compare the predicted value of this latter with its real value.
Figure \ref{fig_2D_superp} shows the galaxies' \hir vs. a selected
colour $g-r$, one of our input features.
The simulated targets are shown with the blue contours
and the predicted values with the green contours. Each column represents 
3 selected setups (see Table \ref{setups}) that only use SDSS magnitudes
during the training whereas each row corresponds to one training model.
The $z$-trained models shown here
(two right columns: zSCmb, zSClr) are at $z=0$.

Overall, the ML-predicted values follow the true values from the
simulation, and show that galaxy colour is anti-correlated with
\hir\ as expected.  The mean trend is always well recovered using any of
the predictors.  However, the scatter in the data is not fully
captured by any of the models: The green contours are always inside
the blue contours.  Different ML algorithms perform differently in
this regard: We see that for DNN, RF \& $k$NN, the two contours are
quite close.  Only looking at the $f$-trained models (left column)
where we train on all the data from $z=0-2$ simultaneously, it is
evident visually that RF maps $g-r$ best, $k$NN comes next followed
by DNN.  For the $z$-trained models where we train individually at
various redshifts, DNN, RF \& $k$NN do similarly well with zSCmb
but the performance of RF is better with zSClr (where we add in the
color indices).  In contrast, SVM, LR and GRAD have difficulty to
capture the scatter in the data, hence their predictions tend
to be more tightly confined around the mean.  While we have shown this
specifically for $g-r$, the results for other colours are similar,
and typically show that RF and DNN perform the best, with kNN not far behind.

\begin{figure}
\includegraphics{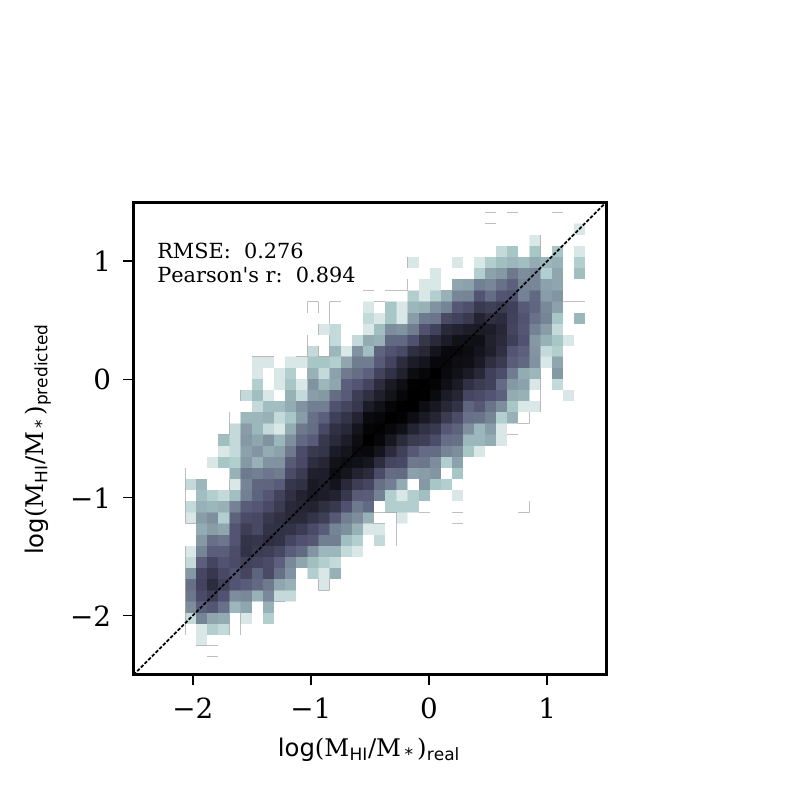}
\caption{2D distribution of the real (x-axis) \textit{vs.} predicted (y-axis) \HI~richness with
the $z=0$-trained DNN model, using the zSCmb training set.}
\label{fig_illus}
\end{figure}

Figure \ref{fig_illus} shows a direct comparison between the real
and the predicted \HI~richness of the galaxies with the DNN models
trained and tested with $z=0$ simulated data. The dashed line shows
the 1:1 line; if the ML algorithm were perfect, all points would
lie along this line.  The correlation is apparent and generally follows the
identity line, indicating that the training performs reasonably well
in the mean.  However, there is a significant scatter, which degrades
the performance on a galaxy-by-galaxy basis.  The best-fit slope
is also not identically unity, so the correlation is not perfect
even in the mean.  We thus would like to quantify our regressors' performance using
the slope and tightness of the correlation.

To quantify the performance of our ML framework, we choose three metrics:
\begin{itemize}
\item The slope of the linear mapping $f: y \rightarrow \hat{y}$, where an ideal mapping
would have a unity slope.
\item {\sc rmse} (Root Mean Squared Error), given by 
\begin{align*} 
\textsc{rmse} = \sqrt{{1\over{N}}\sum^N_{i=1}
                          ({y_{i}  - \hat{y}_{i}})^2 }
\end{align*}
where $y$ and $\hat{y}$ are the real value and the estimate respectively,
gives the average difference between the predicted and the real values.
The square of this metric is also used as
a cost function to be minimized in some methods for regression
($e.g.$ deep neural network, linear regression). The lower the {\sc rmse}
the better the performance of the model is.
\item  Pearson product-moment correlation coefficient (Pearson's $\textbf r$)
which tells how scattered the predictions are compared to the true values. The closer to 1, 
the tighter (or better) the prediction is.
\begin{align*}
\mathrm{Pearson's~r} = {\sum_{i=1}^N (y_{i} - Y)
                                                               (\hat{y}_{i} - \hat{Y})
                                       \over
                                       \sqrt{\sum_{i=1}^N (y_{i} - Y)^2}
                                       \sqrt{\sum_{i=1}^N (\hat{y}_{i} - \hat{Y})^2}
                                       }
\end{align*}
where $Y$ and $\hat{Y}$ are the mean values of  $y_{i}$
and $\hat{y}_{i}$ respectively.
\end{itemize}
In figure \ref{fig_illus}, we get {\sc rmse}$= 0.276$ and Pearson's \textbf{r}$=0.894$
for the particular choice of the DNN regressor and the zSCmb training set; this is
one of our best cases, but RF is actually slightly better.
Previous work by \citet{Zhang-09}, estimating \HI-to-stellar mass ratio using
analytic equation leads to $1\sigma$ scatter $>0.3$,
which shows that our ML approach is more accurate.
\begin{figure*}
\includegraphics{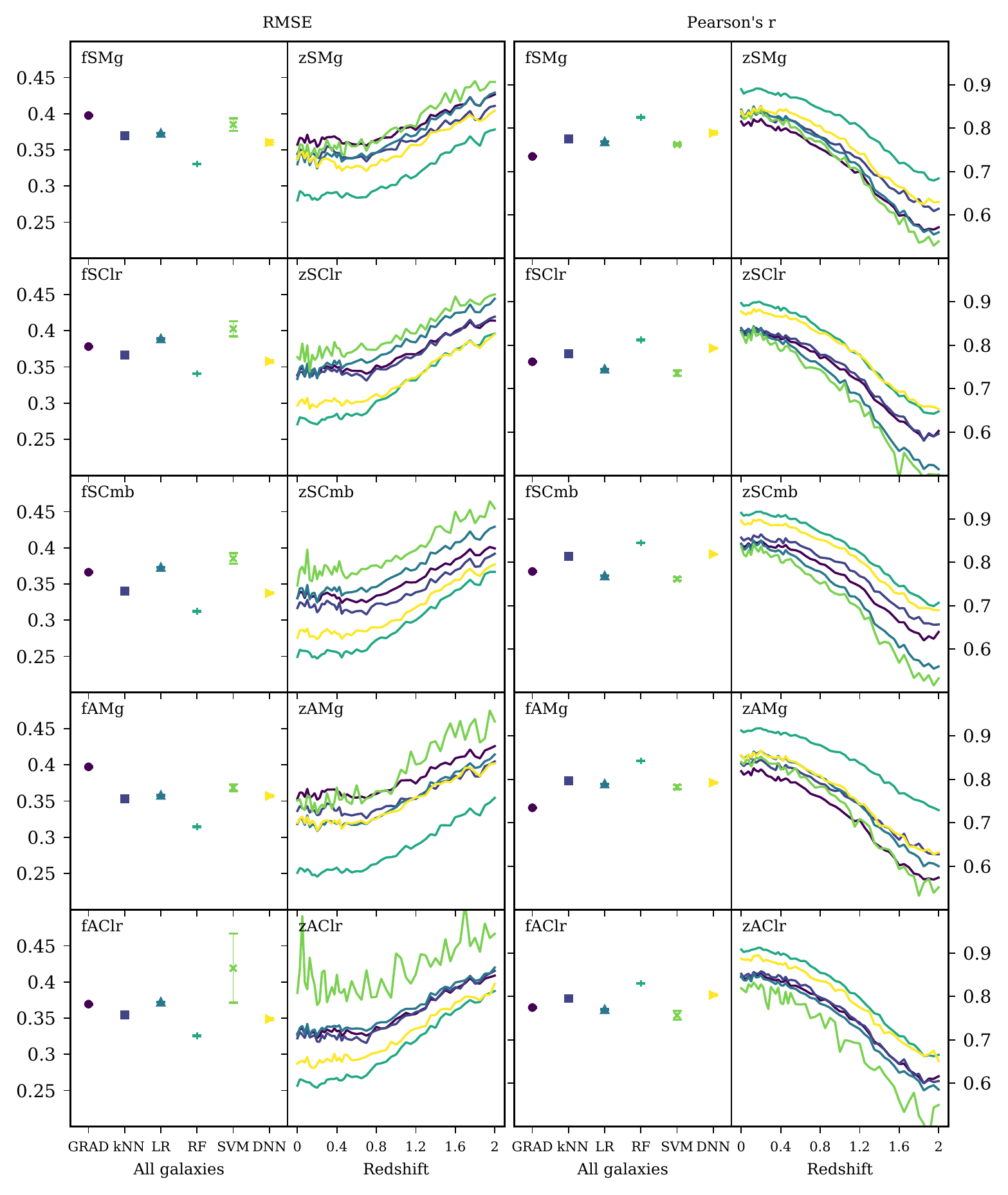}
\caption{Root mean square errors RMSE and Pearson product-moment
correlation coefficients \textbf{r} are shown on the 2 columns from the left and right,
respectively. Models perform better if they show lower RMSE and higher \textbf{r}.
The first on the left shows a mapping for all the galaxies, and the second
for galaxies at different redshifts. The dots and lines are color coded
by the training models we use. Each rows show different results for different
setups. The {\sc rmse} values are shown on the left y-axes and the \textbf{r} values
on the right y-axes.}
\label{fig_correlationall}
\end{figure*}

Figure \ref{fig_correlationall} shows the performance of the various
models considering each setup in Table \ref{setups} using {\sc rmse}
and Pearson's $\textbf r$ coefficient.  The 2 columns from the left
are the {\sc rmse} and the 2 columns from the right are Pearson's
$\textbf r$.  Each row corresponds to the results from different
features used in the training.  The name of the setup is shown on
the top left of each panel.  Different results from different
learning techniques are presented with the color coded lines (with
distinctive markers).  In the following subsections we discuss how
well our various regressors perform when varying the training set
and the training method. 

\subsection{Dependence on redshift}

Examining the leftmost column in Figure~\ref{fig_correlationall},
these are the {\sc rmse}'s for various ML
algorithms when training on the entire data set from $z=0-2$ without
any redshift information ($f$-training).  The results bear out the trends noted
in Figure~\ref{fig_2D_superp}: The RF method generally does the
best (lowest {\sc rmse}) for any of the input data sets, while DNN and
kNN follow, and then the remaining methods.  The RF values are still
typically above 0.3, with the lowest values for the fSCmb (SDSS
colours, magnitudes, and environment) and perhaps marginal improvement
in fAMg which adds the near-IR photometry.

The third column shows the corresponding Pearson's \textbf{r} values.  The
basic story is the same, that RF provides the best prediction, with
values of \textbf{r} $\approx 0.85$ in the best cases, with others down to
\textbf{r} $\approx 0.75$.  The predictions from the aggregate dataset clearly
contain significant information, but are perhaps not as optimal
as one might get from including some redshift information.

The second and fourth columns show the result of training and testing
at individual redshifts ($z$-training).  It is clear that from $z\sim 0-0.5$,
the $z$-training performs better than the aggregate ($f$) training, with
lower {\sc rmse} around 0.25 in the best-case RF models (zSCmb and zAMg).  
The other ML algorithms are clearly poorer than RF, although DNN does
reasonably well in the zSCmb case.  Similarly, the fourth column showing
the Pearson's \textbf{r}  also is very good at $z=0-0.5$, and here DNN in
many cases does nearly as well as RF.

Beyond $z>0.5$, all the regressors show degrading performance, with
increasing {\sc rmse} and decreasing \textbf{r}.  This increase in {\sc rmse}
likely owes to the fact that at high-$z$, all galaxies are more
\HI~rich (\hir$>10^{-2}$) \citet{Rafieferantsoa-15}, with fewer and
fewer quenched galaxies with very low \hir.  Because the intrinsic
\hir\ vs. mass (and other properties) thus becomes fairly flat, it
becomes increasingly difficult for the ML to pick out the correct
\hir\ based on other galaxy properties as would be reflected in the
photometry.  This is likely an intrinsic limitation of this method,
owing to the evolution of \HI\ in galaxies.

Redshift information can be obtained observationally, amongst other
methods, from photometry or spectroscopy. The latter is still easier
to retrieve than direct \HI\ data, while the former typically
obtains redshift errors of a few percent, which is still good
enough to ascribe a training redshift.  It is clear from the above
results that redshift information is useful to improve the predictions.
Even out to $z\sim 1$, the limit of currently planned surveys, the
predictions do not degrade greatly, it is only at $z>1$ that they
become worse than the aggregate case.  Hence from here on we
will primarily discuss the $z$-training results.

\subsection{Dependence on input features}

The different rows in Fig. \ref{fig_correlationall} show the impact
of varying the input features into the ML framework.  As we have
seen, RF generally performs the best followed by DNN.  GRAD, $k$NN,
LR and  SVM perform similarly poorly regardless of our setups (their
{\sc rmse}'s$\simeq 0.34$), with perhaps GRAD performing the worst.
For this reason, unless otherwise stated, we are only going to discuss RF and DNN in what
follows.

At $z=0$, using only SDSS magnitudes results in relatively poor
performance, with {\sc rmse }$\approx 0.3$ for RF and 0.35 for DNN
and others.  For RF, using either \verb'color indices' instead of magnitudes
(zSCls) or in addition to magnitudes (zSCmb) , or including additional
magnitudes into the near-IR (zAMg) improves this significantly,
with {\sc rmse } as low as 0.25 and \textbf{r}$>0.9$.  Thus it appears
that providing colour information directly into the ML algorithm
helps it determine a better mapping than only providing the magnitudes,
even though in principle the magnitudes contain all the colour
information.  Also, providing additional near-IR bands seems to
be advantageous.

For DNN, the story is slightly different.  Again, only SDSS bands
has the worst performance, but here, including the near-IR data
does not improve things as much as providing \verb'color indices',
and particularly providing both \verb'color indices' and magnitudes
together (zSCmb), which achieves a performance approaching that of
RF.

The redshift dependence of {\sc rmse} and \textbf{r}  is similar among all
these combinations of input datasets.  The overall message is that
providing more bands is better, which is unsurprising, but also that
it is preferable to provide the colours directly rather than the
magnitudes given the choice.  In many cases, it is possible via
SED fitting to obtain a galaxy colour that has uncertainties that
are smaller than would be obtained by just subtracting magnitudes,
so this may be a more valuable input for ML predictions.

\subsection{The slope of the mean relation}

\begin{figure*}
\includegraphics{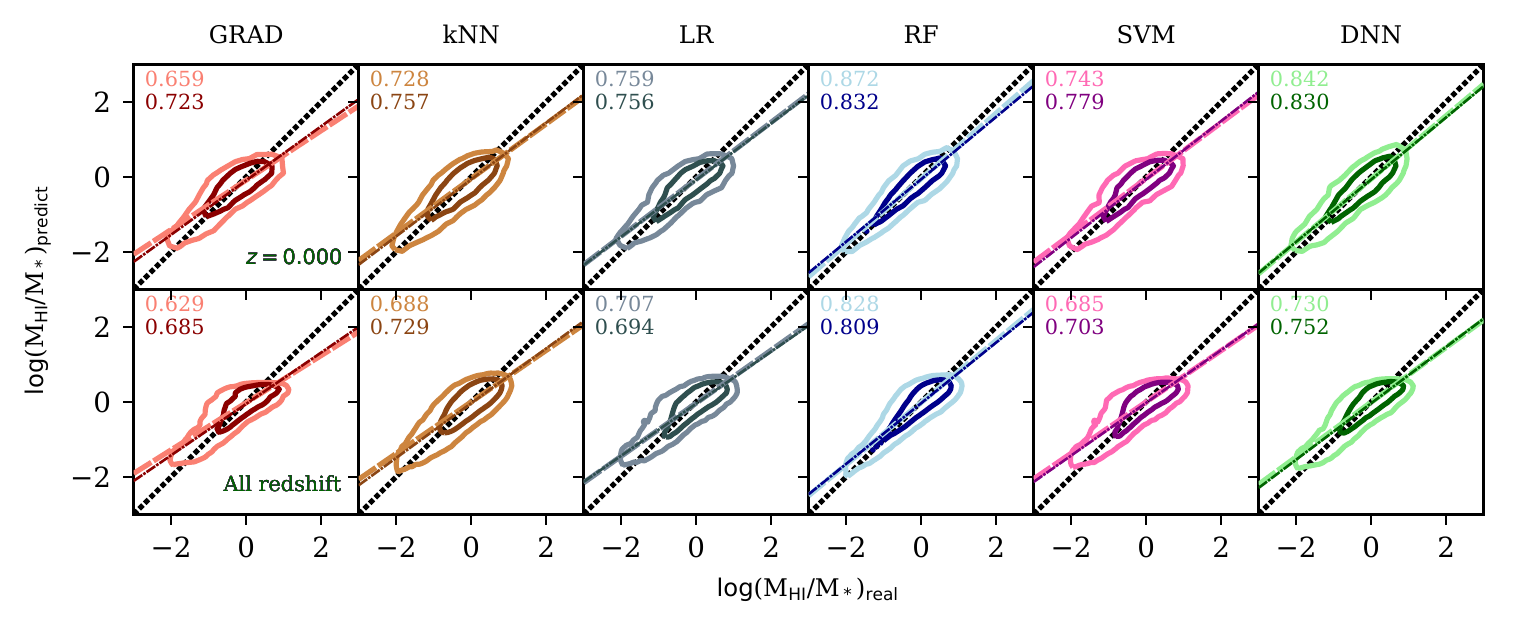}
\caption{2D representations of our real (x-axes) \textit{vs.} predicted (y-axes)
values of \HI~richness. Upper panels show for different models at $z=0.0$,
whereas the lower panels show for all redshift combined.
We only show the results from our \{f,z\}SCmb features.
The numbers with dark (light) colors
on the left top corners show the slopes of the linear fit
of the $1\sigma$ ($2\sigma$) subsample.}
\label{fig_2D_richness}
\end{figure*}

In Figure \ref{fig_2D_richness} we show linear fittings for the
correlation between real (x-axis) and predicted (y-axis) values for
\hir. The top pannels are for the $z$-trained models at $z=0$ and
the lower panels for $f$-trained models.  Each column corresponds
to a given regressor as labeled on top.  In each panel, the dark
(light) lines represents the $1\sigma$ ($2\sigma$) contours between
the targets and the predictions. The numbers on the top right are
the slopes of the linear fits (color coded) for the two contours.
The thick dashed line shows the 1-to-1 relation, which would be the
perfect prediction.  We only show the SDSS combined setup (zSCmb)
here, \textit{i.e.} the features are SDSS magnitudes+\verb'color indices' 
+$v_{gal}$+$\Sigma_3$, but the results from other setups
are similar.

We can see that $f$-trained (lower panels) models tend to have
slopes further away from unity compared to those from the $z$-trained
ones. This confirms what we found previously with {\sc rmse} and
Pearson's {\bf r}, that at low redshifts, training on the smaller
but more homogeneous sample at a given $z$ provides a better prediction
than training on a larger sample that conflates all the redshifts.

Among regressors, again we see that RF and DNN have slopes that are
closest to unity, and thus perform better.  All other methods have
best-fit slopes below 0.8.  However, all the slopes are $<1$, which
indicates an under-prediction of the \HI~richness for \HI~rich
galaxies and over-prediction for \HI-poor galaxies.  This reflects
the fact that, as seen in Figure~\ref{fig_2D_superp}, the true scatter in the
\hir\ around the mean is not fully reproduced in the predictions,
such that all the regressors tend to fit galaxies closer to the mean.
Hence at the lowest \hir, they tend to fit slightly higher values,
while at the highest \hir, they tend to fit slightly lower values,
resulting in a sub-unity slope: akin to an Eddington bias.
The slope thus partly reflects a
measure of how well the scatter around the mean is predicted. 
The fact that RF and DNN have the best slopes just quantifies the
qualitative impression from Figure~\ref{fig_2D_superp} that these
regressors reproduce the extent of the scatter most closely.

\begin{figure}
\includegraphics{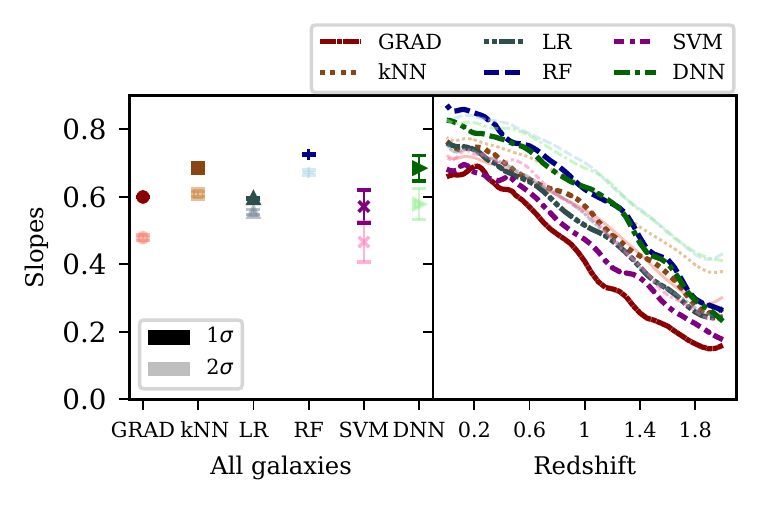}
\caption{Slopes of the linear fit (y-axes) of the relationship between the predictions
and the real \HI~richness of our simulated galaxies.
The dark color (or thick lines) show the fit for the $1\sigma$ sample around
the maximum and the light color (or thin lines) for $2\sigma$.
The left (x-axis showing the names of the models) is similar to what is shown 
in Figure \ref{fig_2D_richness} second row,
the right (x-axis showing the redshift values) panel presents the evolution of slopes from our zSCmb features.}
\label{fig_slopes_all}
\end{figure}

Figure \ref{fig_slopes_all} shows the comparison of slope values
for the $f$-trained sample (left panel) and the redshift evolution
of the $z$-trained sample (right panel) among the various regressors.
The left panel effectively just shows a plot depicting the numbers
in the bottom row of Figure \ref{fig_2D_richness}.  Here, RF performs
the best but not so far from DNN (considering the variance among 10
subsamples), and the other models perform somewhat worse.

The right panel extend the values shown in the upper panel of Figure
\ref{fig_2D_richness} to higher redshift.  Dark colors (or/and thick
lines) show the $1\sigma$ slopes and the light colors (or/and thiner
lines) show the $2\sigma$ slopes.  Looking at the $z$-training
results (right panel), it is very clear that the slopes of RF and
DNN are closer to unity than the other models, and that is true
across all redshifts. The $2\sigma$ slopes (light color lines) are
generally better than the $1\sigma$'s, except at the lowest redshifts.
Slopes $<0.5$ implies a weak correlation between the predicted and
the real values of \HI~richness, so Figure \ref{fig_slopes_all}
indicates that all regressors become unreliable beyond $z\ga 1$.

In summary, $k$-NN, RF and DNN methods show better performance as
compared to SVM, GRAD and LR (Figure \ref{fig_2D_superp}).  DNN and
RF tend to perform better when providing galaxy colours as opposed
to photometry, and when providing more bands.  Among our tests, the
best mapping of \HI~richness was achieved with RF at $z=0$ using
optical and near-IR bands, which gave {\sc rmse}'s $\approx 0.25$
and \textbf{r} $>0.9$.  Using all data from $z=0-2$ together did not provide
as a good fit as training at individual redshifts, despite the
smaller samples for the latter.  The evolution of {\sc rmse} or
Pearson's {\bf r} shows a stronger redshift dependence beyond
$z\sim0.5-1$ making the prediction uncertain at higher redshift
($z>1$, see Figure \ref{fig_correlationall}).  Slopes of linear
fits are generally less than unity owing to the fact that the true
scatter is not fully spanned by the prediction; again, RF performs
the best with DNN close behind, and the other regressors significantly
poorer.  All slopes move further from unity with increasing redshift,
once again limiting applicability at $z\ga 1$.

\section{Application to RESOLVE data}\label{S:test}

We now apply and test our ML methodology against real observations
from the RESOLVE data.  This survey provides both photometry and
\hir, so provides an ideal sample to test the efficacy of our
predictions.  There are two ways we will test this:  First, we will
train on the RESOLVE data itself, and predict the RESOLVE data, to
test how well it works in the ideal circumstance of having the
training and testing set be from the same sample.  Second, we will
train on the simulation and predict the RESOLVE data, which is more
like the application envisioned for this technique, to see how much
degradation there is when the training and testing sets are different.
If the simulation was a (statistically) perfect representation of
the RESOLVE data, we would expect the resulting {\sc rmse} and \textbf{r} 
to be similar, but given that we expect some differences, we aim
to quantify the degredation in a real-world situation.

\subsection{Simulated vs observed data}\label{S:sod}

We first describe the RESOLVE data.  We make use of the photometry
data \citep{Eckert-15}  as well as their corresponding \HI-flux
\citep{Stark-16} from the Data Release \Rom {2} of the RESOLVE
survey.  We use the following standard equation
\begin{equation}\label{MHI}
M_\mathrm{HI} = 2.36 \times 10^5 \times D^2 \times F_\mathrm{Total}
\end{equation}
to compute the \HI~mass in $\msolar$, where $D$
is the distance to the galaxy (Mpc) calculated from the apparent and
absolute magnitude in \textbf{r}  band given in the photometry data.
$F_\mathrm{Total}$, provided by the RESOLVE data,
is the total \HI~line flux ($Jy.\kms$) of the galaxy.
The RESOLVE photometric data release\footnote{\url{https://resolve.astro.unc.edu/data/resolve_phot_dr1.txt}} contains
SDSS (\textit{u,g,r,i,z}), 2MASS (\textit{J,H,K}), GALEX (\textit{NUV}) and
UKIDSS (\textit{Y,H,K}) band magnitudes.

One immediate issue when comparing to simulations will be that
stellar population models, initial mass function, etc. used to
obtain $M_*$ from the data (from which we compute \hir) is different
between what we assume in {\sc Loser} versus what RESOLVE assumed
to obtain their $M_*$ values.  Hence it turns out there is a small
offset in $M_*$ that we must first correct.  We do so empirically,
by using our ML framework to predict the $M_*$ from the photometry
in our simulations and from RESOLVE, and then comparing the $M_*$
values.

\begin{figure}
\includegraphics{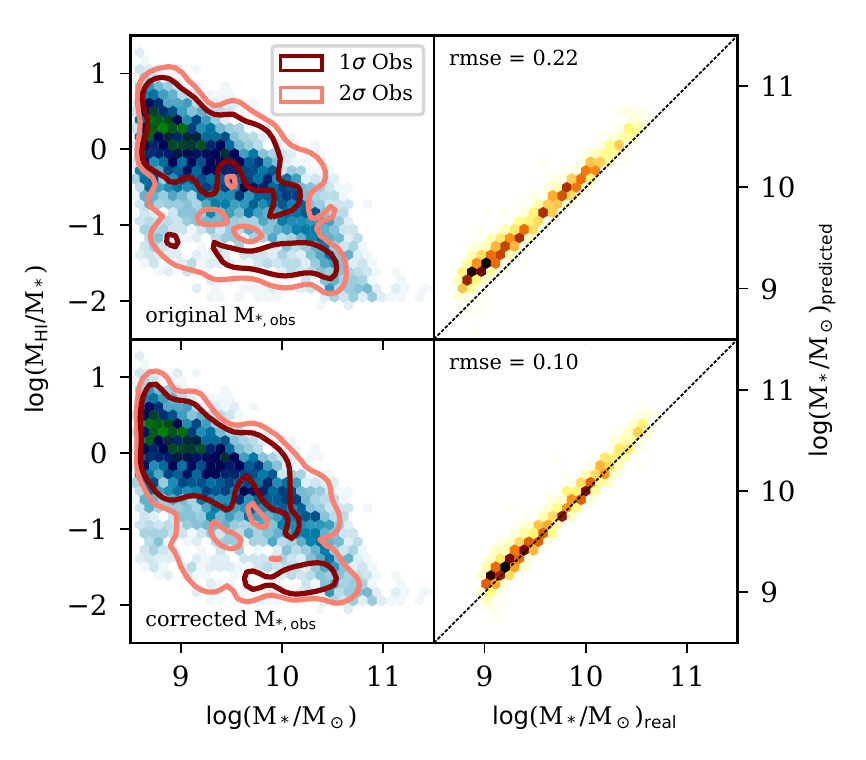}
\caption{
Left panels: The blue-green maps show the distribution of M$_*$ (x-axes) \textit{vs.}
\hir~(y-axes) of the simulated galaxies, while the dark and light red contours
show the 1 \& 2 $\sigma$ distributions of the RESOLVE data.
Right panels: distributions of the real (x-axes) and predicted (y-axes)
galaxy stellar masses of the RESOLVE galaxies. Upper panels show
the distributions prior to the correction
to observed stellar masses as described in the text, and lower panels after correction.
The lack of bimodality in the simulated data ({\it right} panels) as seen
in the data is mainly due to our cut to only include galaxies with \hir$>10^{-2}$.}
\label{fig_correction}
\end{figure}

Figure \ref{fig_correction} (right panels), shows the difference
between the original (top) and the corrected (bottom) M$_*$ values
from RESOLVE.  The original RESOLVE data is offset by $\sim 0.1-0.2$~dex;
this is within the uncertainties of typical $M_*$ determinations from
photometry.  The correction we apply is a linear scaling of the
stellar masses to match with \muf~galaxies, obtained by training
the DNN model with the simulation to predict the stellar mass of
the RESOLVE data, and comparing the result with the real value from
RESOLVE. We repeat the process $10\times$ and take the average of
the linear slopes and the intercepts, to obtain the following
relation: $\log \mathrm{M}_\mathrm{*,corrected} =
0.920\times\log \mathrm{M}_\mathrm{*,original} + 0.924$.
It can be seen that
$M_*$ is predicted very tightly, with a scatter of {\sc rmse}$=0.1$
once the correction is applied.  Prior to the correction, the {\sc
rmse}$=0.22$ relative to the 1-to-1 line, which is dominated by the
offset rather than the scatter itself.  Note that scaling the
simulated stellar masses would give the same results,  but we don't
use this option because we know exactly the stellar mass of the
simulated galaxies.

We can also compare the trend of \hir\ vs. $M_*$ in the simulations
and RESOLVE, which is done in the left panels of Figure
\ref{fig_correction}, before (top) and after (bottom) the $M_*$
correction.  The green-blue distributions on the left panels are
from \muf-galaxies whereas the contours are from the observational
data.  In general, particularly after the correction is applied,
the simulations and observations agree quite well for the bulk
of the galaxies.  A clear trend is seen that lower-$M_*$ galaxies
have higher \HI\ fractions.  The mean trend of the galaxies with
\HI\ is in good agreement between RESOLVE and this simulation,
which confirms the agreement versus other data sets shown
in \citet{Rafieferantsoa-15}.  This indicates that \muf\ provides
a generally viable model to predict observed \HI\ from photometry.

There is a notable difference that the observational data shows a
bimodal distribution that is not seen in the simulated data.  This
is because we have explicitly ignored galaxies from \muf\ with
\hir$<0.01$.  In \muf, we have many galaxies with no \HI, while in
the observations there is a distribution of low-\hir\ values.  We
will leave more careful modeling of these low-\hir\ objects for
future work, but we note that the bimodality is going to degrade
our results since the ML is unlikely to effectively predict galaxies
with \hir\ approaching $\sim 0.01$.

\begin{figure*}
\includegraphics{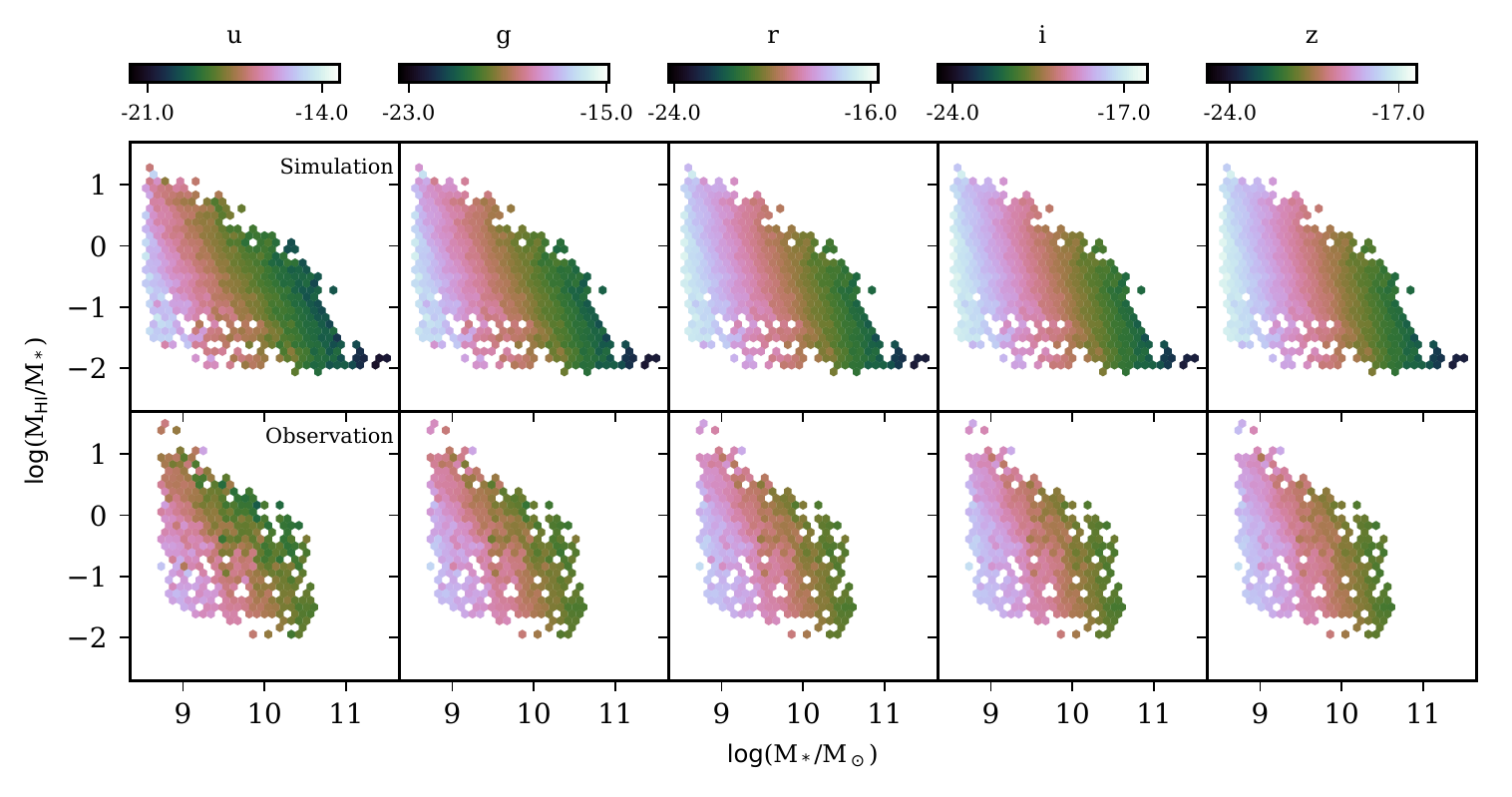}
\caption{The x-axes and y-axes represent the stellar masses and \HI~richness
of the galaxies, respectively. Similar to bottom left panel of Figure \ref{fig_correction},
but showing the mean magnitudes in each pixel for
the SDSS passbands (columns). The top and bottom panels are for simulated
and observational data respectively. The agreement between the two data
is noticeable and the range of the observational data are well included
in that of the simulated ones.}
\label{fig_mag_compare}
\end{figure*}

We also check if the range of magnitudes between the RESOLVE and
\muf\ are in broad accordance.  Figure \ref{fig_mag_compare} shows
the same distributions as in Figure \ref{fig_correction} lower left
panel, except that now the colours in each hexagonal bin represent
the mean magnitudes of the galaxies in that bin.  We show \textit{ugriz}
magnitudes for illustration but we get similar results for other
bands.  Each column represents one band. Upper and lower panels are
for simulated and observational data respectively. We can clearly
see that the trends are consistent.  Note that apart from SDSS
magnitudes, we also use \textit{NUV, J, H} and $K_s$ magnitudes in the training.
We however point that including all those bands decreases the size
of the sample due to missing data in each band.  The RESOLVE data
contain $2159$ galaxies with SDSS magnitudes.  When accounting for
\textit{NUV, J, H} and $K_s$ we end up with only $1017$ galaxies.

\subsection{Training on and predicting RESOLVE data}

We first consider the case where we train the regressors using one
subset of the RESOLVE data and test them using the other subset
(the one which was not used for the training).  Due to the relatively
small sample in hand, we only use 10\% of the data for testing.
This case can be considered optimal in the sense that the training
and testing sets are drawn from (different parts of) the same sample,
so there are no systematic differences.

The right panel on Figure~\ref{fig_SIMOBS} shows our prediction
using the test sets. Judging by the contours, it is clear that all
the presented models here perform reasonably well, \textit{i.e.}  the
distribution of the real vs predicted values lie along the identity
line, and the predicted values (y-axis) covers all the range of the
real values for all regressors.  Comparing regressors, GRAD with
{\sc rmse}$=0.28$ performs best followed closely by RF, $k$-NN and
lastly DNN with {\sc rmse}$=0.44$.  Now the trend is reversed such
that DNN, which was among the best in the previous scenario becomes
the worst in this case.  DNN's typically require larger training
samples to properly constrain the large number of layers, so it is
likely its poor performance owes to the small sample of
RESOLVE galaxies.

This result already has interesting real-world applications.  For
instance, it can be used to populate SDSS galaxies that lack \HI\
data or have poorly constrained \HI\ measurements with \hir\ values.
This would allow a reasonable characterisation of what their \HI\
content would be if RESOLVE had been able to observe them.
Alternatively, one could use the larger ALFALFA sample cross-matched
with SDSS data.  By training the regressors on ALFALFA data along
with their corresponding SDSS photometric data, we can predict the
\HI~content of galaxies that have SDSS photometric data but do not
have ALFALFA counterparts.  The key is that we need a single
photometric sample, for which we have a training set of \HI\ data.
In such a case, our method appears to be able to predict \hir\ to
$<0.3$~dex scatter, which is competitive with and typically better
than previously proposed fitting formulae.

\subsection{Training on \muf\ and predicting RESOLVE}

A more general application would be where we have no or very limited
\HI\ training data, and only photometric data.  This might be the
case at $z\sim 0.3-1$, where the \HI\ data is almost nonexistent now and
even future surveys will provide only a sparse sampling of the most
\HI-massive objects.  In this case, we would like to be able to use
the simulations to provide the training set.  Naturally, this
introduces more uncertainties and assumptions, because the simulations
build in a specific physical model which likely is not exactly
correct, and does not reproduce the real \HI\ population in all
its details.  To test how much more uncertain the predictions would be,
we can attempt this using RESOLVE where we {\it know} what the
correct answer is, and see how well the simulation recovers it
relative to the case in the previous section where we used
RESOLVE itself to train.

In order to mitigate the effects of those uncertainties, one
must carefully mimic the input features of the simulated data
to encompass those from the observational data as discussed in the
previous section.  Given that \muf~reproduces several observables
that are usually used as benchmark for simulation models, such as
stellar mass function, \HI~mass function, specific star formation
rate function, etc. \citep{Dave-16, Dave-17a, Dave-17b}, we feel
confident that it provides a state of the art approach to making
predictions for upcoming surveys such as LADUMA or MIGHTEE, i.e.
using simulated data for training the algorithms and applying it
to available observational photometric data.

\begin{figure}
\includegraphics{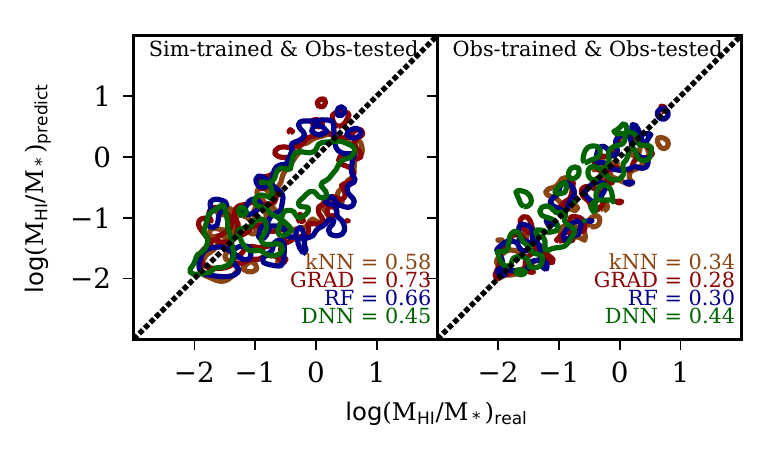}
\caption{Predictions of the observed \HI~richness (y-axes) using different
mapping algorithms (colour coded lines). Left panel shows the results when
the algorithms are trained with simulated data. Right panel shows the results when
training the algorithms with observational data. The contours correspond
to 1$\sigma$ distribution.}
\label{fig_SIMOBS}
\end{figure}

Figure \ref{fig_SIMOBS}, left panel, shows the \HI~richness prediction
of our four best models, training the regressors with the simulation
data and predicted the \HI~richness of the RESOLVE data. The contours
show the distributions of the RESOLVE \HI~richness (x-axis) vs the
predicted \HI~richness (y-axis) from the models.  The numbers on
the bottom right of each panel show the {\sc rmse} of each model.

Overall, the predictions still lie along the one-to-one relation,
indicating that using the simulations to train still provides an
adequate prediction in the mean.  However, the {\sc rmse} values
are much higher here than in the right panel.  This clearly shows
that the simulated sample does not fully mimic the details of the
observed sample.  Given the discrepancies between simulation and
observation, implying differences of the underlying distributions
of the two samples, this is not surprising.

$k$-NN, GRAD and RF now all have {\sc rmse} values above 0.5, which
is fairly poor.  They estimate with larger scatter and a noticeable
offset towards lower \HI~richness values, where the contours get
as far as 1 dex below the 1:1 line at log$_{10}$(\hir) $\sim0$.

Rather remarkably, DNN (green contour) now performs the best in
this case, with {\sc rmse} = 0.45 and predictions extending to the
lowest values ($-2\leq$) following the 1:1 line.  Although DNN was
outperformed in Figure \ref{fig_correlationall} using only simulated
data for training and testing, we can clearly see here that its
performance shines in a more difficult scenario, where now the
training sample is much larger but the data is more complex.  Indeed,
the {\sc rmse} for DNN hardly changed at all when using the RESOLVE
or \muf\ data to train, though this probably arises from the larger
training sample offsetting the less homogeneous testing sample.
Our results suggest that in this real-world application, DNN can
learn better from the simulated data than simpler regressors.

From those two approaches, {\it left} and {\it right} panels of
Figure \ref{fig_SIMOBS}, we can see that DNN presents robust predictions
regardless of the training setups. It is able to learn important features
from the simulation and translate those into the observed data. kNN, GRAD
and surprisingly RF are less efficient in doing so. The latter only performs best
when the training and testing samples are drawn from the same main sample.\\

To summarize, we have shown that training on a subset of observational
data can yield a reasonably tight prediction for a testing set taken
from the same data.  This provides a way to populate photometric
surveys in scenarios where a sizeable \HI\ training set is available,
such as RESOLVE or ALFALFA.  Training the models with simulated
data and predicting the observational targets has higher uncertainties,
but is still feasible.  One has to carefully model the input features
from the simulation to mimic those in the data, which requires
further work.  While RF has generally been the best choice for
regressor in more homogeneous training-test set situations (simulation-simulation
and observation-observation), when applying the simulation training
to observational data, the performance of DNN clearly outshined the
others.

\section{Discussion}\label{weakness}

Extraction of information in  a given set of data is a challenge
in all models.  Although RF and DNN are our best models, they still
have difficulty in extracting all the necessary information,
particularly DNN.  That being said, attaining an accuracy of \textbf{r} $>85\%$
is a non trivial success for both of regressors.  In
our training for the DNN, we make sure that the loss function stays
unchanged for several training steps to make sure the network
learns as much information as it needs  but not as much as it might
overfit the training data and loose the important  information
necessary for the prediction.  It may be possible to tune this
better. 

\begin{figure}
\includegraphics{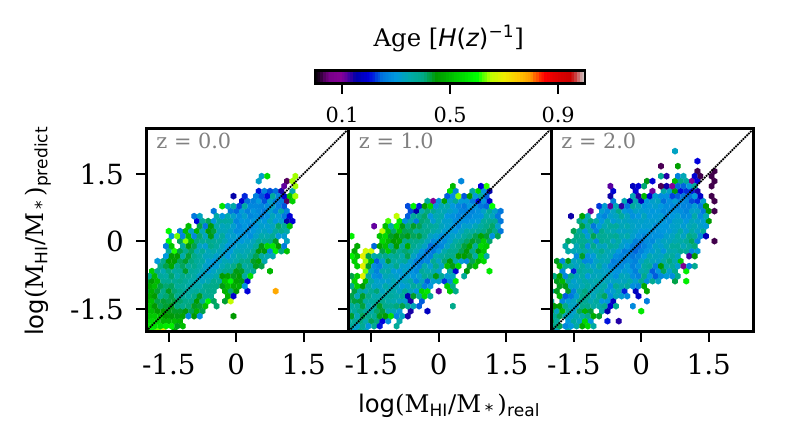}
\caption{Mean galaxy age for each pixel in the distribution of the real (x-axes) and
predicted (y-axes) \HI~richness of the simulated galaxies. This result is from
the DNN-trained model. Different panels show for different redshift.
We use the age of the galaxies at the given redshifts
(shown on the top left corner in each panel).}
\label{fig_Age}
\end{figure}

It is possible that photometric surveys can yield other information
such as the age, star formation rate, and (from a group catalog)
halo masses, albeit with some uncertainties.  It is interesting to
ask whether providing such information would improve predictions.
However, we find that this is unlikely to be the case.  We illustrate
this for the mean stellar age in Figure \ref{fig_Age}.   Here we
show the distribution of the galaxies based on their real \HI~richness
and the predicted values from the DNN model, with the colour of
each hexagonal bin showing the mean age of galaxies falling in that
bin (in unit of the Hubble time at the given redshift).
Different panel show different redshifts: {\it left, center,
right} for $z=\{0,1,2\}$ respectively. We can see that for a given
\HI~richness value we cannot see any age gradient in the predicted
values, and it remains the case up to $z=2$. We interpret this to
mean the ML model has learned about the age of the galaxies even
though that information was not explicitly given in the training
set.  The same situation happens with the specific star formation
rates and the halo mass of the galaxies.  This is the case for all
of our ML models.  Hence providing such information, which introduces
further uncertainties from their estimation, is unlikely to be helpful.

Then we might ask why do some models perform better than others?
We believe that the design of the models themselves may lead to different
mapping of the input-output, thus, to improved results depending
on the data.  Changing the layer structures in DNN or optimising
the tree size (or the number of base estimators) in RF might alleviate
certain issues we encountered in our training. We are currently
analyzing such possibility and might improve our model in that
direction in upcoming work.  Also DNN may particularly benefit from 
a larger simulation training sample with more dynamic range than
available in \muf.

\begin{figure}
\includegraphics{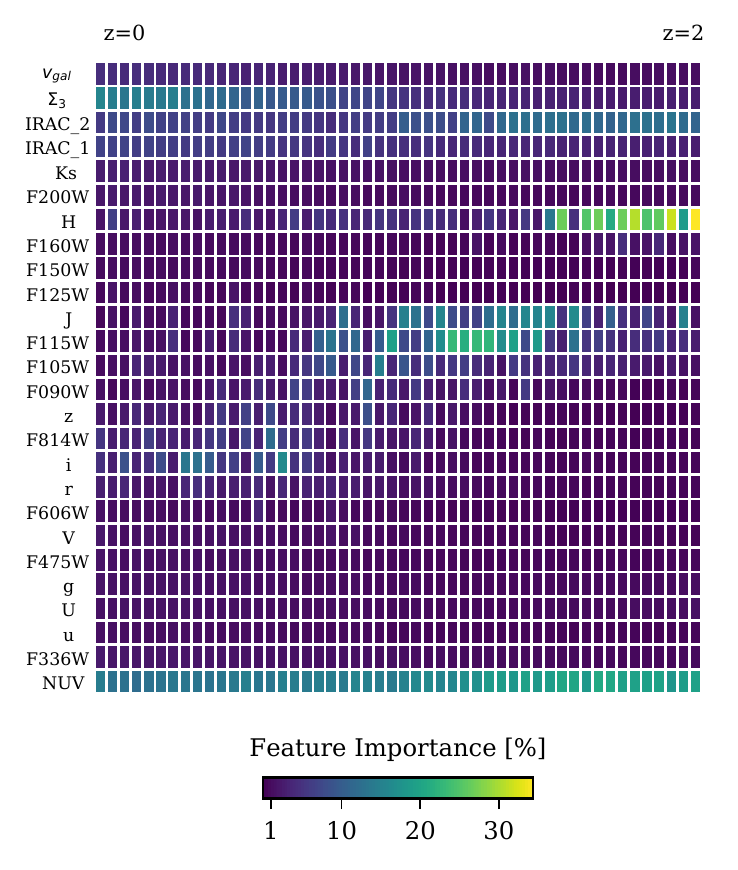}
\hspace*{1mm}
\includegraphics{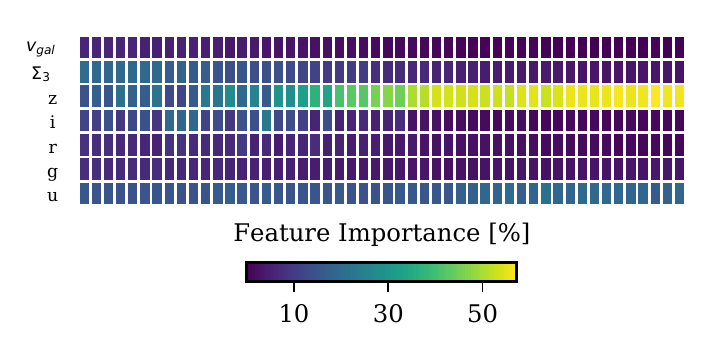}
\caption{Evolution of the importance of the input features from the RF training.
Each row represents one band with the filter name on the left, except for the
1\textsuperscript{st} (2\textsuperscript{nd}) row which show for the line
of sight velocity (3\textsuperscript{rd} nearest neighbour) feature. The bands
from the bottom to the top are with increasing peak wavelengths.
Left to the right shows the feature importance from $z=0$ to $z=2$.}
\label{fig_importance}
\end{figure}

One useful feature of RF is that it provides an estimate of the
importance level of the input parameters, based on the rate of
incidence that a given parameter is utilised in the decision trees.
We show in Figure \ref{fig_importance} the importance of parameters
from RF training. The upper subfigure shows the result when using
all the available magnitudes in from our simulation whereas the
lower subfigure represents the result when only using the SDSS
magnitudes.  The 1\textsuperscript{st} (2\textsuperscript{nd}) row
in each subfigure show the importance of the line of sight velocity
$v_{gal}$ (3\textsuperscript{rd} nearest neighbour $\Sigma_3$) from
$z=0$ (left) to $z=2$ (right).  The remaining rows show for bandpass
filters (names on the left) with a wide range of peak wavelenghts
from 2309\AA~(bottom row) increasing to 44630\AA.  It is interesting
to see that $\Sigma_3$ becomes increasingly important only at later
epochs. The line of sight peculiar velocities $v_{gal}$ do not add
value to the training, which is unsurprising since it is not obvious
why the \HI\ content should care about peculiar velocity (except perhaps
through correlations of peculiar velocities and the large-scale
potential well); this
in a sense serves as a sanity check that our method is not finding
physically implausible relationships.  In the upper
subfigure, the IRAC channels have some importance at higher redshift,
particular IRAC $4.5\mu m$ while $3.6\mu m$ is less important. The
H-band magnitude is very important at high redshift but contributes
much less at low redshift.  The importance of magnitudes between i
($6250\AA$) and J ($12500\AA$) bands move from low to higher peak
wavelengths towards higher redshift.  NUV magnitudes seem to exhibit
relatively high importance at all redshift bins, highlighting the
connection between \HI\ and the gas that fuels star formation and
hence UV light.

In the lower subfigure with a more restricted input set, $z$ magnitude
is very important at higher redshift but becomes less although still
important at $z=0$, whereas the importance of $i$ magnitude increases
towards the present day.  The value that $u$ magnitude adds to the
accuracy of the prediction seems to be relatively constant at all
redshifts, following NUV in the upper subfigure. 

On the whole what the two panels in Figure~\ref{fig_importance}
tell us is that given the features available in the data, the feature
importance in principle allows one to select only a set of the most
important ones in order to achieve a given accuracy. This, amongst
other methods like Principal Component Analysis (PCA), is of a great
value especially when reducing the dimensionality that might not
be avoidable due to a limited computing power or when the dimension
is as big as the size of the data (\textit{i.e.} number of features
is as large as the number of examples for the training).  Also, the
importance levels could be helpful in survey design, if a particular
photometric band is more useful it might be regarded as higher
priority to obtain.  However, one must be aware that in many cases,
RF importance levels do not truly reflect the necessity of a given
data, in the sense that sometimes RF says a particular input is important,
but the information from that input is actually encoded in the 
other inputs, so that removing it does not have as detrimental
effect as one might think.  Properly assessing the importance level
would involve re-training the entire data set removing each input
in turn, to assess the increase in {\sc rmse}.  Nonetheless, RF
importance level can at least provide a guide in this process.

\section{Conclusion}\label{conclusion}

We have investigated estimating the \HI~richness of galaxies based
on their optical and near-IR survey properties, in particular SDSS
$\{u,g,r,i,z\}$, Johnson $\{U,V\}$ and 2MASS $\{J,H,K_s\}$, line
of sight velocities, and environmental measures, using machine
learning (ML).  For our analysis, the training data have been generated
from the \muf~simulation. We have tested various machine learning
regressors including random forests and deep neural networks.  We
considered various input feature combinations, including only SDSS
magnitudes and environmental properties, using galaxy colours instead
of and in addition to magnitudes, and including 2MASS and Johnson
magnitudes.  We trained each model to predict \hir\ based on an
aggregate of all simulated galaxies at $z=0-2$ ($f$-training), and
in 50 individual redshift bins ($z$-training).  As an example application,
we applied this framework to the RESOLVE galaxy survey catalog with
\HI\ and photometric data.  To measure and compare the performance
of each method, we used {\sc rmse}, Pearson correlation coefficient
\textbf{r} , and the correlation slope.

We summarize our main findings as follows:
\begin{itemize}

\item By using 75\% of the \muf\ data for training and testing on
the remaining quarter, we find that all ML methods are able to
approximately recover \hir\ from galaxy photometry.  The accuracy
depends both on the input data set and the ML algorithm.  Generally,
random forests (RF) provides the best performance at $z=0$, i.e.
lowest {\sc rmse} $\approx 0.25$, highest \textbf{r} $ \approx 0.9$, and slope
closest to unity, with deep neural network (DNN) close behind.

\item At $z\la 1$, it is advantageous to do the ML training at a
given redshift rather than aggregating all redshifts.  The smaller
number of galaxies available for training in the former is outweighed
by the conflating of evolutionary trends when aggregating. 
The {\sc rmse} of all ML algorithms increases with redshift, with
commensurately lowered \textbf{r}  and a best-fit slope diverging from
unity, though the effect is mild out to $z\sim 0.5$.  Predictions
at higher redshifts are more challenging owing to reduced trend in
\hir\ among high-$z$ galaxies, since most galaxies at $z\ga 1$ have
similar \hir\ prior to significant populations of quenched galaxies
arising.

\item Providing more input training data results in better predictive
power, unsurprisingly.  Using only SDSS data results in {\sc
rmse}$\approx 0.3$ for RF at $z=0$, while either including 2MASS
data or training on both colours and magnitudes yields a more optimal
{\sc rmse}.  DNN has in the best case similar performance, but it is
more strongly dependent on the selected input features. 

\item All the regressors tend to under-predict the high \HI~richness
and over-predict the low \HI~richness, as shown by the slope ($<
1$) of the linear fits between the targets and the predictions.
This owes to the regressors being unable to fully capture the scatter
in the \hir\ values at \textit{e.g.} a given colour, instead tending to push
the \hir\ towards the mean.  This raises the value of low \hir\
objects and lowers it for high \hir\ objects, resulting in a sub-unity
slope.  The under-prediction of the high \HI~richness is more severe
at high redshift (Figure \ref{fig_slopes_all}).

\item By training our ML framework on a subset of the RESOLVE data
and testing it on the remainder, we showed that it is possible to
predict \hir\ with {\sc rmse} $\approx 0.3$, which is comparable or
better than what is obtained with scaling laws; RF again performs
among the best, though GRAD is slightly better.  When training on
\muf\ and testing on RESOLVE, we find the best regressor is DNN,
but the predictions are significantly degraded with {\sc rmse}$\approx
0.45$, likely owing to subtle mismatches between simulation predictions
and analysis procedures and those from RESOLVE.  While the scatter
is substantial, the mean trend remains well-matched, showing that
the ML algorithm introduces only mild systematic biases, and thus is
still valuable for statistical survey applications.

\end{itemize}

We have shown through this study that it is clearly possible to
estimate the \HI~ richness of a galaxy by relying only on the
information from photometric magnitudes. We considered various
magnitudes from different surveys like SDSS, Johnson and 2MASS in
this work, but including other bands is doable.  The broadly
successful test on RESOLVE data suggests that the estimation of
\HI~gas at higher redshift (being $z\leq1$) using the methods
presented here, even with the lack of testing data, is sensible.
With the advent of future surveys such as LADUMA and MIGHTEE, our
ML framework constitutes an important new tool to aid studies of
neutral hydrogen and galaxy evolution.

For our analysis, we have only selected galaxies that are observable
in \HI, with a threshold of \hir$>10^{-2}$.  This raises a key
question:"{\it Would a model still generalize well if one also
included the \HI-depleted galaxies in the dataset for the training?}".
There are two ways to address this question:

\begin{itemize}
\item We can simply add the \HI-deficient galaxies in the dataset and redo
the fitting procedure prescribed in this work, although from the standpoint
of observations, predicting the \HI~richness of a \HI-depleted or gas-starved
galaxy is not really meaningful. 
\item The more elegant approach would be to first use ML to classify galaxies based
on their observable features
whether they are \HI~deficient or not, then only estimate its \HI~richness
(based on the same features) in the case it would potentially contain observable
\HI. Of course, the minimal value of observed \HI~ can be a free parameters
in our model but in reality that should depend on the telescope capabilities.
\end{itemize}

Future work will discuss these solutions, provide more tailored
predictions for upcoming surveys, utilise larger training samples
that could particularly help improve DNN results, and make this
tool available to the community.

\section*{Acknowledgements}
The authors thank S. Hassan, B. 
Moews and G. I. G. J\'ozsa for helpful conversations and guidance.
MR and RD acknowledge support from the South African Research Chairs
Initiative and the South African National Research Foundation.
MR acknowledges financial support from the Square Kilometre Array
post-graduate bursary program.
SA acknowledges financial support from the Square Kilometre Array.
The \muf~simulations were run on the Pumbaa astrophysics
computing cluster hosted at the University of the Western Cape,
which was generously funded by UWC's Office of the Deputy Vice
Chancellor. MR uses additional computing resources from
the Max Planck Computing \& Data Facility (\url{http://www.mpcdf.mpg.de}).




\bibliographystyle{mnras}
\bibliography{paper_bib}


\label{lastpage}
\end{document}